\documentclass[twocolumn]{aastex631}

\usepackage{graphicx}
\usepackage{subfigure}
\usepackage{longtable}
\usepackage{xcolor}
\usepackage{soul}
\newcommand{\tess}{\textit{TESS}}

\begin{document}

\newcommand{\jt}[1]{\textcolor{blue}{#1}}

\shorttitle{\tess\ sub/lr giant asteroseismology}
\shortauthors{Grusnis et al.}

\graphicspath{{./}{figures/}}

\title{\tess\ Subgiant and Lower Red Giant Asteroseismology in the Continuous Viewing Zones}
\author{Sophia Grusnis}
\affiliation{University of Florida
Gainesville, FL 32611, USA}
\affiliation{California State University,
Los Angeles, CA 90032, USA}

\author{Jamie Tayar}
\affiliation{University of Florida
Gainesville, FL 32611, USA}

\author{Diego Godoy-Rivera}
\affiliation{Instituto de Astrofísica de Canarias (IAC), C/Vía Lactea, s/n, E-38200 La Laguna, Tenerife, Spain}
\affiliation{Universidad de La Laguna (ULL), Departamento de Astrofísica, E-38206 La Laguna, Tenerife, Spain}

\correspondingauthor{Sophia Grusnis}
\email{sgrusni@calstatela.edu}

\begin{abstract}

Asteroseismology, the study of stellar oscillations, and stellar modeling both offer profound insights into the fundamental properties and evolution of stars. With \texttt{pySYD}, a new open-source Python package, we were able to constrain the asteroseismic global parameters, $\nu_{\text{max}}$ and $\Delta\nu$, for 82 solar-like oscillating subgiant and lower red giant stars, filling in the region between the {\it Kepler} dwarfs and giants. Using asteroseismic scaling relations, we were able to compute seismic masses, radii, and surface gravities for our entire sample with average errors of 0.21 $M_{\bigodot}$, 0.27 $R_{\bigodot}$, and 0.06 dex respectively. Using 4 stellar modeling grids we determine and compare stellar ages for our sample. We find that our age distribution from stellar modeling is consistent with other local star samples. We find small consistent offsets from model predictions across our regime, but offsets were worse at higher gravities (log(g) $\geq$ 3.5 dex), suggesting the need for better calibration. Finally, we discuss our sample in the context of galactic archaeology and show how ages like these could be used to identify and study binary system evolution and galactic evolution in the future. All in all, we show that asteroseismology can be successfully performed with \tess\ data and can continue to make an impact on our understanding of stellar physics and galactic archaeology.

\end{abstract}


\section{Introduction} \label{sec:intro}

Our proficiency in stellar modeling is key to understanding many facets in astronomy. However, stellar models are based on many free parameters which need calibration and experimentation. In order to check, validate, and improve the models, we need a sample of stars with known mass, radius, temperature, and composition. Previous efforts have used eclipsing binaries \citep{2017A&A...608A..62H} or stars in open clusters \citep{2010RSPTA.368..755K, 2024A&A...681A..13F}, but asteroseismology allows us to do similar work for the more numerous single field stars.

Asteroseismology, the study of stellar oscillations, provides precise and accurate stellar parameters which we use to check these models \citep{2010ApJ...725.2176Q, 2017MNRAS.467.1433R, 2019MNRAS.484..771R}. To test these models, we have asteroseismically characterized some of the best subgiant and lower red giant solar-like oscillators in the \tess\ \citep{2015JATIS...1a4003R} Continuous Viewing Zones (CVZs) at both the North Ecliptic Pole and South Ecliptic Pole. Solar-like oscillators are defined as stars whose pressure-mode oscillations are excited by convection, like our Sun \citep{2010aste.book.....A}. These types of oscillators include dwarfs cooler than about 6300 K, subgiant stars, and giant branch stars. For solar-like oscillators, each star displays a characteristic pattern of oscillations with a Gaussian envelope around the frequency at maximum power, $\nu_{\text{max}}$, with a regular frequency separation, $\Delta\nu$. These global parameters are used to find seismic masses and radii using a scaling off of the solar values \citep{1995A&A...293...87K}.

The {\it Kepler} \citep{2010Sci...327..977B} and \tess\ \citep{2015JATIS...1a4003R} satellites have delivered precise light curves for hundreds of thousands of stars. Significant work has been done using the {\it Kepler} data to test our understanding of giants and dwarfs \citep{2010ApJ...713L.176B, 2011ASPC..448..167G, 2024arXiv240117557S}. However, the subgiant and lower giant stars were understudied in {\it Kepler}. For subgiant stars, oscillations tend to occur at frequencies around 200-400 $\mu$Hz and therefore require high precision at a cadence faster than 30 minutes. During the {\it Kepler} mission, the high cadence slots were generally used for dwarf and main sequence (MS) turnoff stars, so only a handful of subgiant and lower giant stars were included in \citep{2012ApJ...756...19D, 2014A&A...564A..27D, 2020MNRAS.495.2363L}.

These stars are particularly interesting because on the subgiant branch, luminosity is strongly correlated with mass and age, allowing a possible avenue for the estimation of large numbers of precise ages \citep{2017ApJS..232....2X, 2020MNRAS.495.3431L, 2022Natur.603..599X}. However this is also a region where models can differ significantly \citep{2012A&A...548L...1D, 2020MNRAS.499.2445H}. Therefore, using this technique requires careful calibration of the models to ensure they represent a correct mapping between luminosity and age.

Since these stars were understudied in {\it Kepler}, we turn to \tess\ as a possible avenue for this calibration, and in particular to a set of stars in the \tess\ CVZs, regions having an ecliptic latitude of about 78$^{\circ}$ above and below the ecliptic plane, which were targeted for 2 minute cadence observation with more than 1 year of data for each subgiant and lower giant star. We know from previous work like \citet{2019ApJS..241...12S, 2019AJ....157..245H, 2023A&A...669A..67H} and \citet{2024ApJS..271...17Z} that this is enough data to do asteroseismology with our sample and compare our findings with the predictions of stellar models. Our stellar sample could also be useful in the future to identify mixed modes which would give valuable information on the internal stellar structure.

In this paper, we aim to give a first look at the available sample of targets in both the northern and southern CVZs. We give initial estimates of stellar masses, radii, metallicities, surface gravities, and approximate ages for our entire sample. Although there are more sophisticated techniques that can be used to find age estimates for these stars including fitting of individual modes, mixed modes, or detailed modeling, for the purpose of this paper, we use the simplified methods more commonly used for large samples of red giants in order to get a more general sense of our sample.

\section{Methods} \label{sec:methods}

\subsection{Sample Selection} \label{subsec:sample}

The stars in our sample were chosen for their potential for detailed seismic characterization in the region between dwarfs and giants that was understudied in {\it Kepler}. We wanted stars that have many months of data, which was most common in the \tess\ CVZs. We targeted stars that had the potential to be cool enough to have mixed modes detected \citep{2014A&A...572L...5M}, i.e. an effective temperature less than or equal to 5600K, but were small enough that they would be unlikely to be well studied in the initial 30-minute cadence full frame images (log(g) $\geq$ 3.3 dex). We used the Asteroseismic Target List \citep{2019ApJS..241...12S}, and the detection probabilities therein to prioritize stars; in the South we required stars in our sample to have a detection probability greater than 50\% from {\it Gaia} DR1 \citep{2016A&A...595A...4L}, in the North we required stars to have a detection probability greater than 50\% using {\it Gaia} DR2 \citep{2018A&A...616A...2L}. We also prioritized stars that already had high resolution spectroscopic data at the time of targeting. In total we present 82 solar-like oscillating subgiant and lower red giant stars.

\subsection{Light Curve Creation} \label{subsec:LC}

\begin{figure*}[t!]
    \centering
    \begin{subfigure}[]
        \centering
    \includegraphics[width=1.0\textwidth,clip=true]{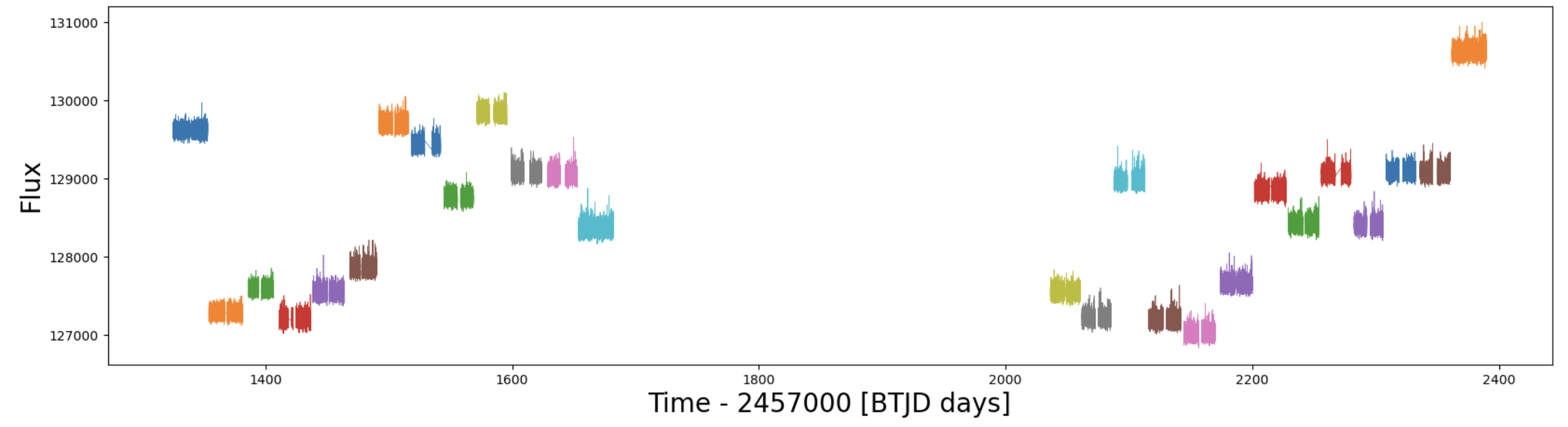}
    \label{subfig:all}
    \end{subfigure}

    \begin{subfigure}[]
        \centering
    \includegraphics[width=1.0\textwidth,clip=true]{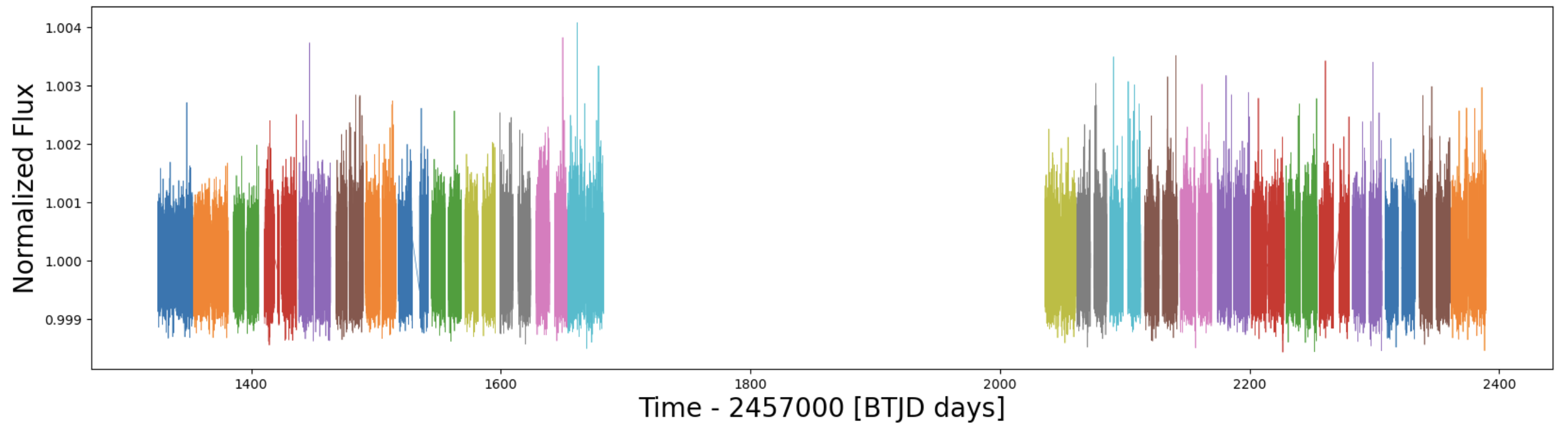}
    \label{subfig:LC}
    \end{subfigure}
    \caption{(\textbf{TOP}): Unedited time series for TIC 38602419. Before any normalizing, we can see the differences in each sector, including sectors 1-13 and 27-39. We show the totality of the two years of data that \tess\ provides, including the obvious discontinuities between sectors. Each sector is distinguished by a different color.
    (\textbf{BOTTOM}): The final normalized light curve for TIC 38602419. The sector colors are consistent between the two plots.}
    \label{fig:LK}
\end{figure*}

To measure oscillations we require a star’s light curve and power spectrum. Given the \tess\ observing strategy and our focus on the CVZs, most of our stars have about 24 months of data with a gap of at least 1 year in between the observed sectors. To collect these light curves, we made use of \texttt{Lightkurve}, a Python package for {\it Kepler} and \tess\ data analysis \citep{2018ascl.soft12013L}. Before any corrections, we used \texttt{Lightkurve} (v2.4.2) to download all the available sectors to look for odd variations and large time gaps (Figure \ref{fig:LK}a). We excluded sector 50 from our analysis because of the extra noise for many stars, corresponding to the time frame 2664.3-2693.0 [2457000 BTJD days]. We then used \texttt{Lightkurve} to normalize the light curves calculated by dividing the flux by the median flux (Figure \ref{fig:LK}b). We chose to keep all of the intersector and intrasector gaps in the \tess\ data collection so as to not alter our inferred seismic results \citep{2014A&A...568A..10G, 2022RNAAS...6..202B}. For completeness, we also ran all of our data with the large time gaps taken out and the subsequent times shifted up. There was a median change of 0.0027 $\mu$Hz for $\nu_{\text{max}}$ and 0.0073 $\mu$Hz for $\Delta\nu$. In Table \ref{tab:table1} we list our $\nu_{\text{max}}$ and $\Delta\nu$ that were found without time shifting the light curves. 

\subsection{Asteroseismic Analysis} \label{subsec:analysis}

\begin{figure}[ht!]
\plotone{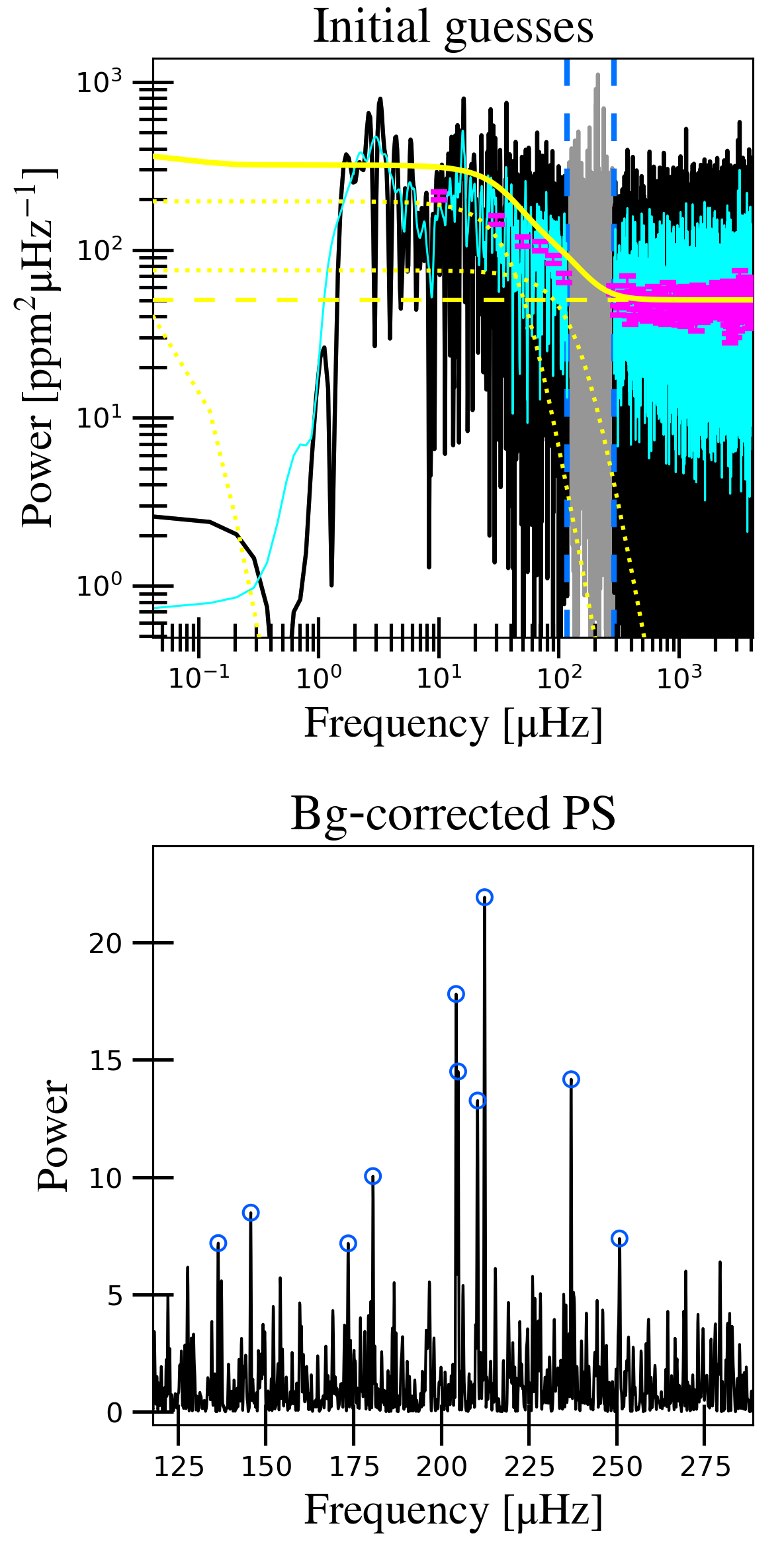}
\caption{Output charts from \texttt{pySYD}. \textbf{(TOP)}: The power spectrum showing the area around $\nu_{\text{max}}$ (shaded region). We see a power excess caused by stellar oscillations (gray vertical band). The black shows the original power spectrum given by \texttt{pySYD} based on our given light curve. The cyan shows a smoothed power spectrum and the pink shows a heavily smoothed power spectrum. The heavily smoothed power spectrum is used as an initial background fit to search for oscillations. We smoothed aggressively in the convective damping timescale to better see the star's oscillations and this resulted in low power at lower frequencies.
\textbf{(BOTTOM)}: The corrected power spectrum with all smoothing removed showing a definite power excess in the same region.  \label{fig:9grid}}
\end{figure}

Once we had our light curves, we needed an asteroseismic analysis pipeline to estimate $\nu_{\text{max}}$ and $\Delta\nu$. For this purpose, we chose to use \texttt{pySYD} \citep{2022JOSS....7.3331C} a new open-source Python implementation of the effective and well-validated Sydney code \citep{2009CoAst.160...74H}. It is designed for automated asteroseismic analysis of solar-like oscillators, primarily focusing on extracting global oscillation parameters from stellar light curves. With \texttt{pySYD}, we are able to analyze frequency spectra which simplifies the process of characterizing stellar oscillations. In addition to the automated analysis, \texttt{pySYD} has a feature to display figures of every processing step along the way. Because of this, we were able to visually inspect our results. Figure \ref{fig:9grid} shows a column from the complete \texttt{pySYD} final display. From the initial plotted power spectrum to the final echelle diagram, we can see that \texttt{pySYD} identified the oscillations of this star (TIC 25156036) and was able to calculate and present an estimate for the $\nu_{\text{max}}$ and $\Delta\nu$ values.

In the analysis, we tried varying a few of the \texttt{pySYD} parameters, i.e. the box filter width, the number of peaks in the autocorrelation function, and smoothing for the background-corrected power spectrum, to ensure we were receiving the clearest output figures. The box filter width is the width of the smoothing window, in units of $\mu$Hz, that is applied to the power spectrum data. Smaller widths, ~1-2 $\mu$Hz, preserve the finer details in the power spectrum but retain some noise, medium widths, ~5-10 $\mu$Hz, gives a more balanced relationship to the smoothing and detail visualization, and larger widths, ~10-20 $\mu$Hz, highlight global trends in the power spectrum, but remove individual peaks. The autocorrelation function (ACF) is important in detecting the $\Delta\nu$ value between consecutive radial modes. Decreasing the number of peaks in the ACF can be useful when the only expected signal is a dominant $\Delta\nu$ signal. However, fewer peaks may only capture the strongest periodicity and miss important secondary features. Increasing the number of peaks in the ACF can detect more periodic spacings but might introduce noise and pick up on false correlations from the noise fluctuations. The background corrected power spectrum is the spectrum after the removal of some instrumental noise and background signal, leaving the oscillation signal. A decrease in smoothing can be helpful to preserve sharp peaks, which then detects more features, but less smoothing can also amplify the noise and make $\nu_{\text{max}}$ and $\Delta\nu$ interpretation more difficult. An increase in smoothing reduces the noise and enhances the global oscillation signal, but can blur smaller features and separations. These three parameters can change the look of the power spectrum and ACF by a lot and that is why small manual changes can help identify what the best output figure looks like. The default box width, 20 $\mu$Hz, and number of peaks in the ACF, 5, worked best for us, so we stayed with these values. The default smoothing of the background corrected power spectrum, 2.5 $\mu$Hz, was over-smoothing our power spectrum, so we changed it to 0.0 $\mu$Hz, and found that output to be more reliable.

\begin{figure*}[t!]
    \begin{subfigure}[]
        \centering
    \includegraphics[width=0.5\textwidth,clip=true]{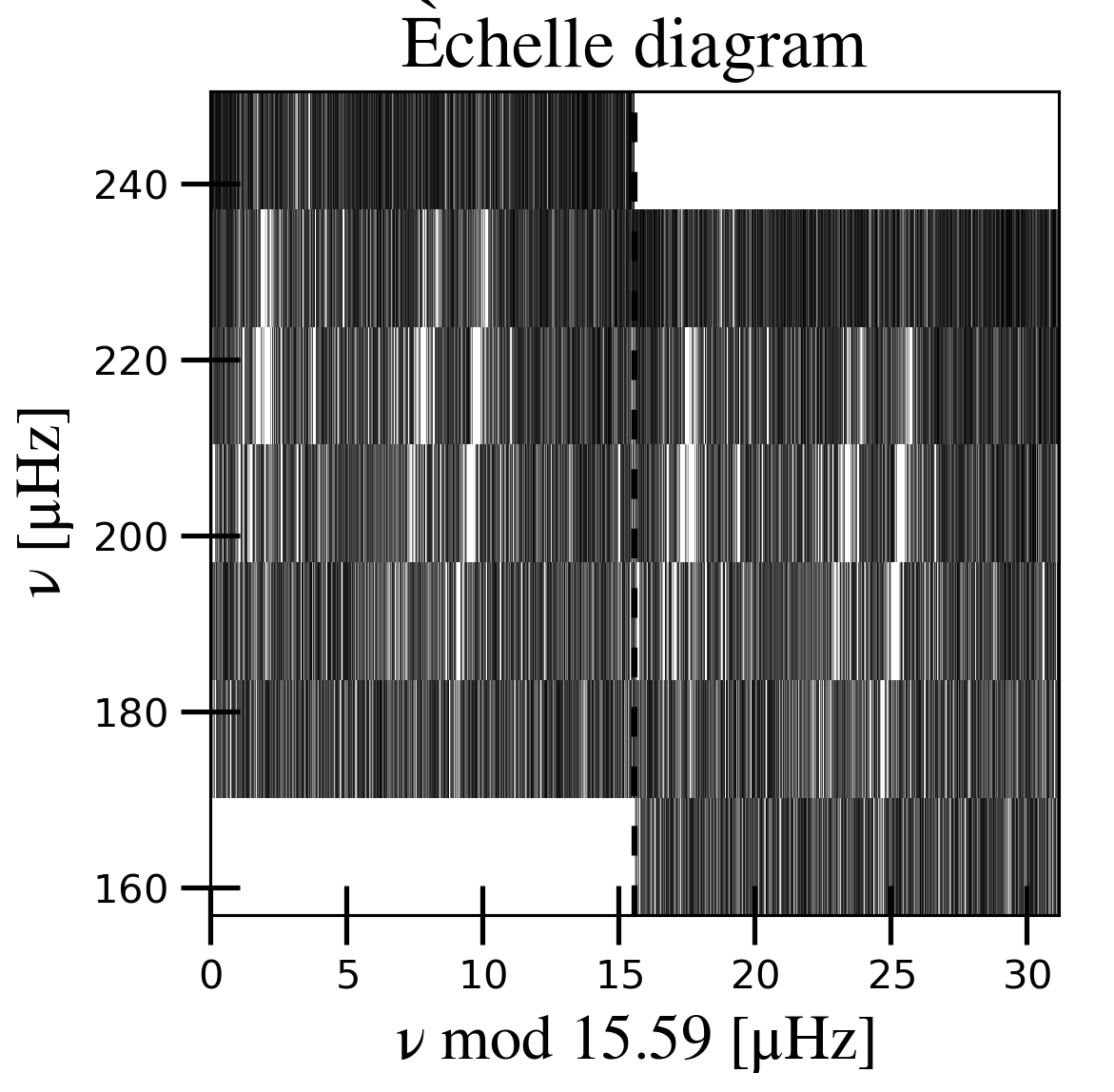}
    \label{subfig:echelle}
    \end{subfigure}
     ~ 
    \begin{subfigure}[]
        \centering
    \includegraphics[width=0.4\textwidth,clip=true, trim=0in 0in 0.4in 0.5in]{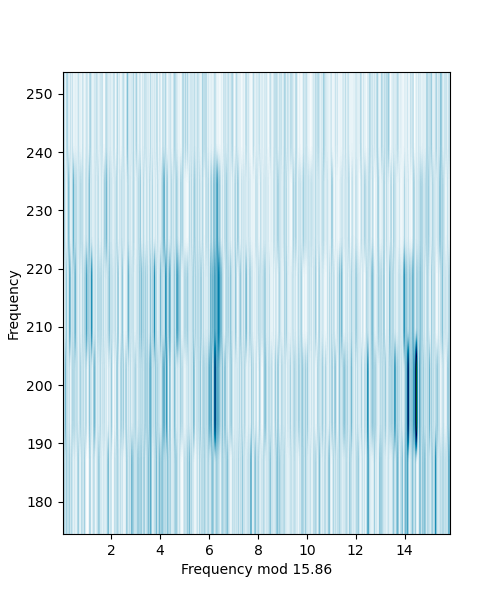}
    \label{subfig:heyechelle}
    \end{subfigure}
    \caption{(\textbf{LEFT}): Echelle diagram given by \texttt{pySYD} for TIC 25156036. With a $\Delta\nu$ of 15.59 $\mu$Hz we can see ridges for the $l$=0, 1, and 2 modes. There is a small tilt to the ridges, meaning we need to do fine tuning in HeyEchelle.
    (\textbf{RIGHT}): Echelle diagram given by HeyEchelle for this same star. With a $\Delta\nu$ of 15.86 $\mu$Hz we can see the ridges are lined up better than our original $\Delta\nu$ estimate.}
    \label{fig:echelle}
\end{figure*}

An echelle diagram is a plot of the frequency modulo the $\Delta\nu$ value, against the frequency (Figure \ref{fig:echelle}a). In order to refine our estimates of the large frequency spacing, $\Delta\nu$, we used the echelle package HeyEchelle \citep{daniel_hey_2020_3629933} to make echelle diagrams for each of the stars. With our power spectrum given by \texttt{pySYD} as the input, HeyEchelle models the echelle spectra by stacking the amplitude spectrum by the frequency mod $\Delta\nu$. This package allows fine manual adjustments to the exact $\Delta\nu$ value until the ridges for the $l$=0, 1, and 2 modes line up exactly. Our final $\Delta\nu$ values were taken from \texttt{pySYD}, based on the aforementioned light curves and power spectra, and were further adjusted based on visualizations from HeyEchelle. Although this visual adjustment makes our uncertainties on $\Delta\nu$ harder to constrain directly, we used the resulting uncertainties on log(g)
and radius, to check a combination of $\nu_{\text{max}}$ and $\Delta\nu$, and our results were consistent with our previous uncertainty claims. These claims were based on uncertainties given from \texttt{pySYD} and changes that were made after a visual inspection of the echelle diagram from HeyEchelle. Using HeyEchelle, we were able to try many $\Delta\nu$ values for one star and see how slight changes in our estimated value can improve our result (Figure \ref{fig:echelle}b).

\subsection{Calibrations and Corrections} \label{subsec:cal/corr}

\subsubsection{Scaling} \label{subsec:scaling}
Once we had our finalized list of $\nu_{\text{max}}$ and $\Delta\nu$ for every star, we used these values to infer their seismic masses and radii. We used the modified version of the scaling relations \citep{1995A&A...293...87K}, from \citet{2018ApJS..239...32P}, to find our initial masses, $M_{sc}$, and radii, $R_{sc}$ defined by
\begin{equation} \label{eq:1}
\left(\frac{M_{sc}}{M_{\bigodot}}\right)=\left(\frac{\nu_{\text{max}}}{\nu_{\text{max},\bigodot}}\right)^3\left(\frac{T_{\text{eff}}}{T_{\text{eff},\bigodot}}\right)^{1.5}\left(\frac{\Delta\nu}{\Delta\nu_{\bigodot}}\right)^{-4}
\end{equation}
\begin{equation} \label{eq:2}
\left(\frac{R_{sc}}{R_{\bigodot}}\right)=\left(\frac{\nu_{\text{max}}}{\nu_{\text{max},\bigodot}}\right)\left(\frac{T_{\text{eff}}}{T_{\text{eff},\bigodot}}\right)^{0.5}\left(\frac{\Delta\nu}{\Delta\nu_{\bigodot}}\right)^{-2}.
\end{equation}
The solar values we used are from the SYD pipeline from that same work, 3090 $\mu$Hz for $\nu_{\text{max}, \bigodot}$ and 135.1 $\mu$Hz for $\Delta\nu_{\bigodot}$. Seismic scaling relations are empirical relationships for solar-like oscillators that connect asteroseismic parameters to fundamental stellar properties. While these relations can be helpful in estimating stellar characteristics from their observed oscillation frequencies, their empirical nature and intrinsic physical assumptions about the relationships between the star being studied and the Sun do result in limitations. From \citet{1991ApJ...368..599B} and \citet{1995A&A...293...87K}, we expect $\Delta \nu$ to scale with the sound travel time and therefore the mean density of the star. We also expect $\nu_{\text{max}}$ to scale as the acoustic cut-off frequency, although the exact relationship observed is empirical. These global parameters, $\nu_{\text{max}}$ and $\Delta\nu$, and their scaling relations are used in many analyses of large samples of solar-like oscillators for computing asteroseismic stellar properties, for example, in \citet{2014ApJS..210....1C} and \citet{2014ApJS..215...19P}.

\subsubsection{Corrections}

Because the scaling relations are not expected to be perfect, we carefully examine and correct each component of them. For each star in our sample, we used a temperature from careful SED fitting \citep[$T_{\text{eff}}$ in][]{2021ApJ...915...19G}, and a $T_{\text{eff}, \bigodot}$ of 5777 K. Previous work has shown that asteroseismic pipelines can have significant and consistent offsets from one another \citep{2018ApJS..239...32P} based on properties such as how measurements are done, how backgrounds are handled, and that internal uncertainties from each pipeline may not reliably account for this \citep{2018ApJS..239...32P}. Given that we are using a new combination of data processing and pipeline on this dataset, we thought it prudent to compare our values to previous works. Our uncertainties for $\nu_{\text{max}}$ and $\Delta\nu$ were calculated by taking the difference between our sample and previous literature, \citet{2023A&A...669A..67H} and \citet{2024ApJ...965..171L}, and subtracting the previous literature's uncertainties in quadrature. We find that this compares well to other analyses of these stars (see Figure \ref{fig:nuvsdnu} in Section \ref{sec:validation}).

There are known deviations from the scaling relations \citep{2011ApJ...743..161W}. For our analysis, we estimate the theoretical $\Delta\nu$ correction for each of our stars using ASFGrid \citep{2016ApJ...822...15S, 2022RNAAS...6..168S} to find $f_{\Delta\nu}$. Specifically, we computed the correction using the $\nu_{\text{max}}$, $\Delta\nu$, $T_{\text{eff}}$, and [Fe/H] of each star. This asymptotic correction factor helps to refine the mass and radius estimates. The deviations from the scaling relations arise because of stellar structure effects. For subgiant and lower red giant stars, there is a high constrast in the density between the core and the envelope which is accounted for by this $f_{\Delta\nu}$ correction factor from ASFGrid.

For the metallicity required in the grid, we used the Salaris correction \citep{1993ApJ...414..580S} calculated using a combination of the star’s [M/H] and $\alpha$-element enhancement, [$\alpha$/Fe], from the Apache Point Observatory Galactic Evolution Experiment (APOGEE) spectroscopic survey \citep{2022ApJS..259...35A}, specifically the values from \citet{2021ApJ...915...19G}. This correction is included because enhancing $\alpha$-elements in stars can increase opacity levels, thereby changing convection depths and subsequently altering oscillations. For the stars with no listed [Fe/H] value, we assumed solar metallicity [Fe/H]=0, which is roughly the average of our sample. Our total uncertainty on metallicity was then calculated by propagating forward the uncertainties on [M/H] from \citet{2021ApJ...915...19G}, and [$\alpha$/Fe] from APOGEE. Our $f_{\nu_{\text{max}}}$ correction, 0.9877$\pm 0.003$, was the multiplicative factor necessary to, on average, have our seismically inferred luminosity match the luminosity from {\it Gaia}, with the uncertainty given by the propagated estimate of the scatter \citep{2022ApJ...926..191Z, 2025ApJS..276...69P}. Once we have both of our corrections, we can then use equations 
\begin{equation} \label{eq:3}
M_{cor}=\frac{f_{\nu_{\text{max}}}^3}{{f_{\Delta\nu}}^4}{M_{sc}}
\end{equation}
\begin{equation} \label{eq:4}
R_{cor}=\frac{f_{\nu_{\text{max}}}}{{f_{\Delta\nu}}^2}{R_{sc}}.
\end{equation}
to find our final corrected masses, $M_{cor}$, and radii, $R_{cor}$, (Table \ref{tab:table1}). The median fractional change in mass from $M_{sc}$ to $M_{cor}$ is -0.053 and for radius from $R_{sc}$ to $R_{cor}$ it is -0.021.

\section{Validation} \label{sec:validation}

As described in the previous section, we used a combination of standard seismic analysis tools and visual inspection to identify and measure oscillation signatures. We used the data given by {\it Gaia} DR2 and APOGEE in order to check our seismic radii and surface gravities.

\subsection{Seismic Validation} \label{subsec:seis val}

\begin{figure*}[t!]
    \begin{subfigure}[]
        \centering
    \includegraphics[width=0.45\textwidth,clip=true, trim=0in 0in 0.65in 0.5in]{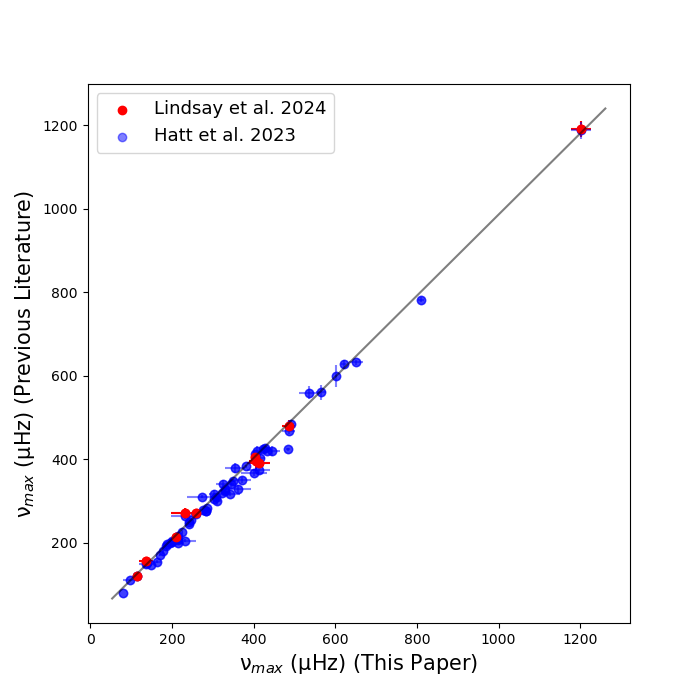}
    \label{subfig:vsnu}
    \end{subfigure}
     ~ 
    \begin{subfigure}[]
        \centering
    \includegraphics[width=0.45\textwidth,clip=true, trim=0in 0in 0.65in 0.5in]{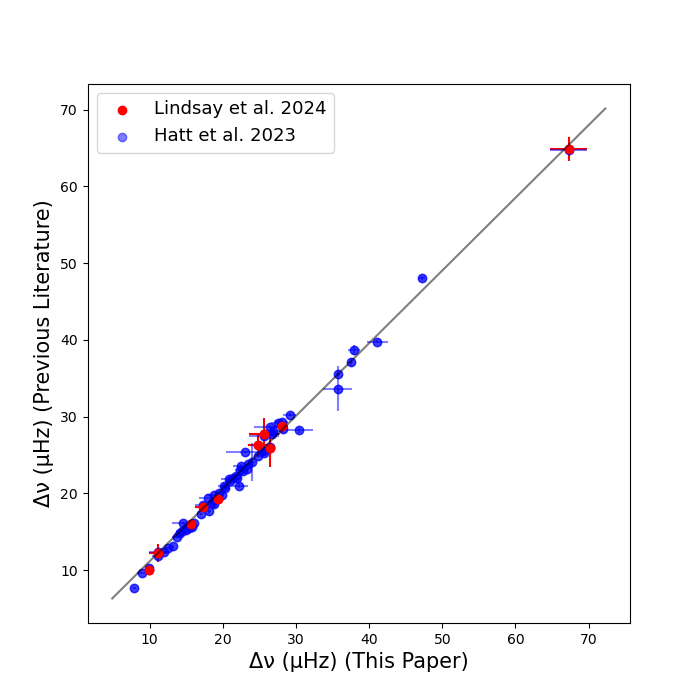}
    \label{subfig:vsdnu}
    \end{subfigure}
    \caption{ 
    Our $\nu_{\text{max}}$ values (left) and $\Delta\nu$ values (right) are consistent with previous work by \citet{2023A&A...669A..67H} (blue) and \citet{2024ApJ...965..171L} (red). We show an overlapping subset of 66 stars out of our total 82 stars. We find an $R^2$ correlation value to \citet{2023A&A...669A..67H} of 0.9918 and to \citet{2024ApJ...965..171L} of 0.9979. We find average offsets of $2.9\%$ in $\nu_{\text{max}}$ and $3.7\%$ in $\Delta\nu$, with no strong trends. Error bars are plotted for all points, but may be smaller than the markers.}
    \label{fig:nudnu}
\end{figure*}

\begin{figure}[ht!]
\includegraphics[width=0.45\textwidth,clip=true, trim=0in 0in 0.65in 0.5in]{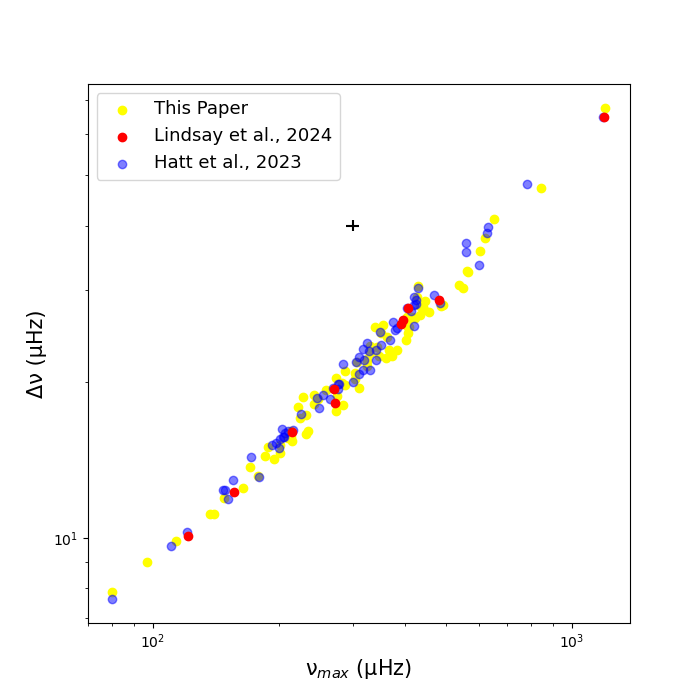}
    \caption{Linear relationship of $\nu_{\text{max}}$ vs $\Delta\nu$. Our unique star sample is shown in yellow compared to the  samples that overlap with \citet{2023A&A...669A..67H} in blue and \citet{2024ApJ...965..171L} in red. Average error bars for the sample are shown in black at the top of the figure.}
    \label{fig:nuvsdnu}
\end{figure}

Previous asteroseismic analyses have included a subset of the stars we study here. For these stars in Figure \ref{fig:nudnu}a and \ref{fig:nudnu}b we show our $\nu_{\text{max}}$ and $\Delta\nu$ values compared directly to the $\nu_{\text{max}}$ and to the $\Delta\nu$ values from \citet{2023A&A...669A..67H} and \citet{2024ApJ...965..171L}. This strong correlation between the seismic observables, $\nu_{\text{max}}$ and $\Delta\nu$, (Figure \ref{fig:nuvsdnu}) is in line with other published literature \citep{2009MNRAS.400L..80S, 2009A&A...506..465H, 2010A&A...517A..22M, 2011A&A...525A.131H, 2011A&A...530A.100H}. The tight agreement with \citet{2023A&A...669A..67H} and \citet{2024ApJ...965..171L}, with average offsets of 0.037\% and 0.029\% respectively, validates our $\nu_{\text{max}}$ and $\Delta\nu$ values. We do see a few offsets and believe these arise from a difference in the methods to obtain the $\nu_{\text{max}}$ and $\Delta\nu$ values. A more careful comparison of multiple methods, like those done for \texttt{Kepler} in \citet{2025ApJS..276...69P} and the \texttt{TESS} SCVZ giants in \citet{2021MNRAS.502.1947M} is left as future work.

\subsection{Classical Validation} \label{subsec:classical}

\begin{figure*}[t!]
    \centering
    \begin{subfigure}[]
        \centering
    \includegraphics[width=0.45\textwidth,clip=true]{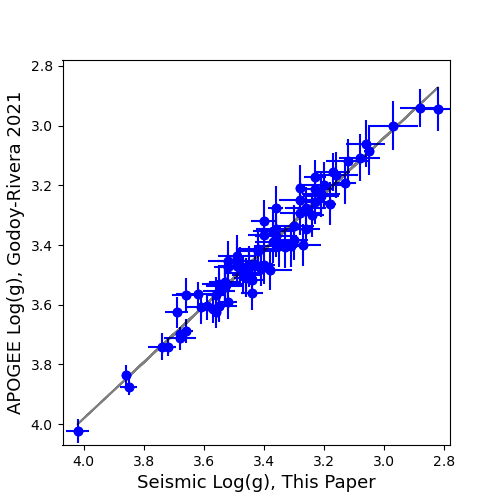}
    \label{subfig:logg}
    \end{subfigure}
    ~ 
    \begin{subfigure}[]
        \centering
    \includegraphics[width=0.45\textwidth,clip=true]{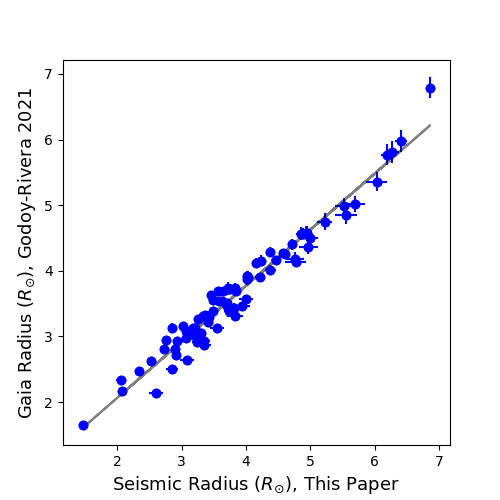}
    \label{subfig:radius}
    \end{subfigure}
    \caption{Our seismic results compare well to spectroscopic surface gravities from APOGEE (left) and radii from {\it Gaia} (right, \citet{2021ApJ...915...19G}. For each plot we show error bars for all points, but may be smaller than the markers.}
    \label{fig:diego_comp}
\end{figure*}

We can also validate our results by direct comparison to results from classical methods. In Figures \ref{fig:diego_comp}a and \ref{fig:diego_comp}b, we show our seismic radii versus the derived SED fit {\it Gaia} radii, and our seismic surface gravities versus APOGEE’s spectroscopic surface gravities as published in \citet{2021ApJ...915...19G}. We find that they are generally well matched, validating our calibration was performed correctly, to the individually calibrated {\it Gaia} and APOGEE values listed in the \citet{2021ApJ...915...19G} paper with average uncertainties slightly lower at 0.27 $R_{\bigodot}$ for radii and 0.06 dex for surface gravities. Seismic radius uncertainties were calculated by propagating forward the uncertainties on the $\nu_{\text{max}}$, $\Delta\nu$, $T_{\text{eff}}$, $f_{\nu_{\text{max}}}$, and $f_{\Delta\nu}$ according to standard error propagation. Surface gravity uncertainties were calculated by taking the $\Delta$log(g) value between our seismic gravities and those from \citet{2021ApJ...915...19G} and subtracting the \citet{2021ApJ...915...19G} uncertainties in quadrature.

\subsection{Detection Probabilities} \label{subsec:prob}

\begin{figure}[ht!]
\includegraphics[width=0.45\textwidth,clip=true]{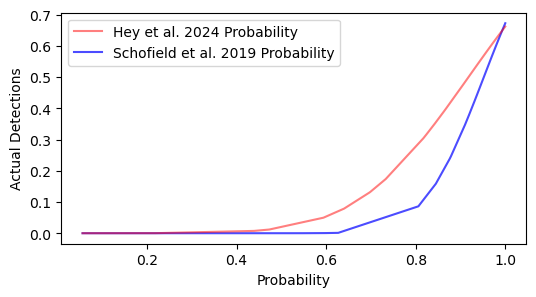}
    \caption{Cumulative distribution function for our solar-like oscillator detections versus the probability of detecting oscillators given by \citet{2019ApJS..241...12S} and \citet{2024arXiv240302489H}. Overall, the ranking of predictions was accurate, but we were less likely to detect oscillations than expected.}
    \label{fig:prob}
\end{figure}

The amplitudes and timescales of solar-like oscillations are expected to scale with the luminosities and temperatures of the stars. Predictions were made in \citet{2019ApJS..241...12S} and \citet{2024arXiv240302489H} about which stars in \tess\ should have detectable oscillations with a probability calculated using astrometric distances, magnitudes in the $I$ and $V$ bands, ($B$-$V$) color, sky position, and effective temperature. We show in Figure \ref{fig:prob} a comparison between these predictions and our detections. In general, we find that the predictions in \citet{2019ApJS..241...12S} and \citet{2024arXiv240302489H} were overly optimistic. While the detection probabilities do correlate with detection rates, we detect oscillators in very few stars with an expected probability less than 70\%. We can see a slight increase in prediction accuracy from the \citet{2024arXiv240302489H} catalog versus the \citet{2019ApJS..241...12S} catalog which could be due to the use of {\it Gaia} DR3 stellar parameters instead of values from {\it Gaia} DR2. It is also possible that with additional sectors or more careful light curve processing one could detect additional oscillators, but in general from the available light curves the noise seems to be higher than expected and therefore detections are less likely.

\begin{figure}[ht!]
\includegraphics[width=0.45\textwidth,clip=true, trim=0.9in 0in 1.1in 0.5in]{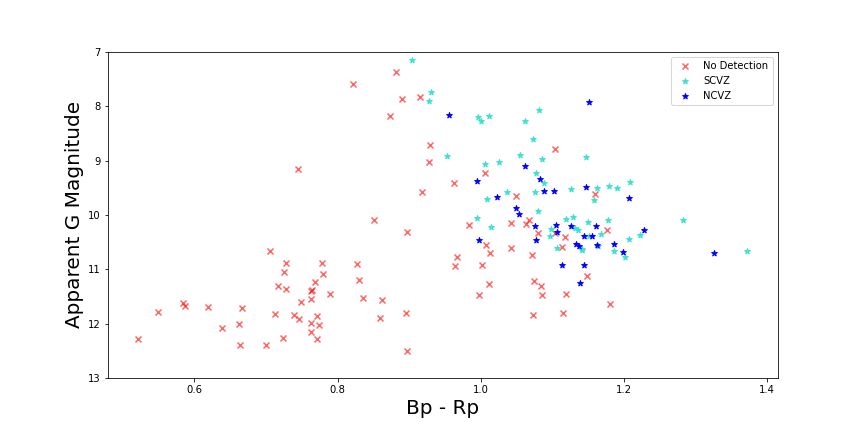}
    \caption{We show that there is a definite trend that brighter and redder stars are more likely to be detected as solar-like oscillators. Our SCVZ stars (light blue) that showed oscillations covered a wider range of colors but were brighter than the NCVZ oscillators (dark blue). The red x markers show stars with no detected oscillators.}
    \label{fig:bprpg}
\end{figure}

When plotting the {\it Gaia} Bp-Rp color versus the {\it Gaia} apparent G magnitude, we found that seismic detections in \tess\ subgiants are color and apparent magnitude dependent (Figure \ref{fig:bprpg}) with detections among early subgiants tending to be rare (i.e., ($BP$-$RP$)$<0.9$ mag, apparent $G$$>11$ mag). With short cadence data, we should be able to detect these oscillations. However, this color-magnitude trend implies that the noise level is too high, and it is possible that we would need the 20-second cadence data, which seems to have less noise \citep{2022AJ....163...79H}, to actually identify these signals.

\begin{figure*}
    \centering
    \includegraphics[width=0.8\textwidth]{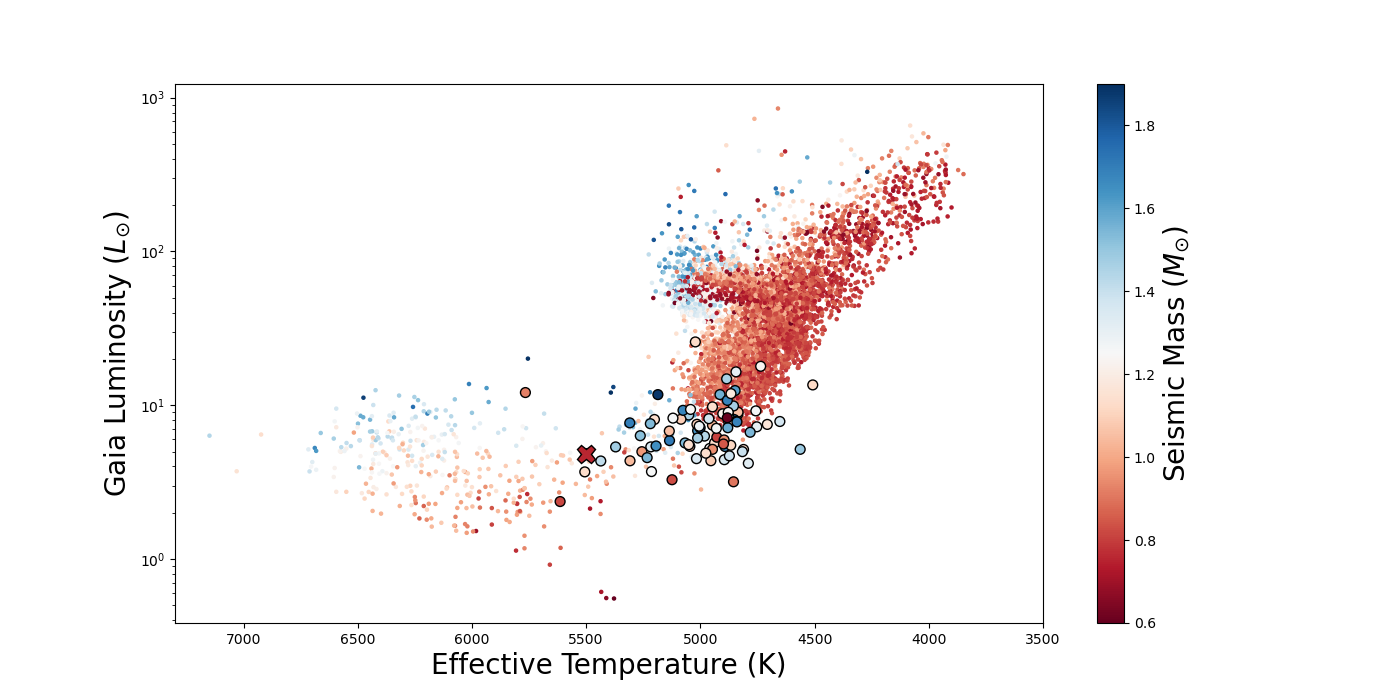}
    \caption{Asteroseismic masses as a function of temperature and luminosity. Our data is outlined with black circles and compared to previous result from {\it Kepler} for dwarfs and subgiants \citep{2017ApJS..233...23S}, and giants \citep{2018ApJS..239...32P}. We also emphasize the location of one unusual star, TIC 366490487 (see Section \ref{subsec:outlier}) with an x marker at a $T_{\text{eff}}$ of 5499 K, and a luminosity of 4.84 $L_{\bigodot}$. In general, our seismic results are consistent with previous efforts, but populate a regime undersampled in {\it Kepler}.}
    \label{fig:HR}
\end{figure*}

\subsection{Sample Distribution} \label{subsec:dist}

In Figure \ref{fig:HR} we compare the luminosities and effective temperatures, derived in \citet{2021ApJ...915...19G}, of the stars we detect oscillations for to those studied in {\it Kepler} \citep{2017ApJS..233...23S, 2018ApJS..239...32P}. In general we find that our stars are more evolved than the well studied dwarfs \citep{2017ApJS..233...23S} and less evolved than the well studied giants \citep{2018ApJS..239...32P} as expected, filling in the region between these two {\it Kepler} samples.

\begin{figure}
\includegraphics[width=0.45\textwidth,clip=true]{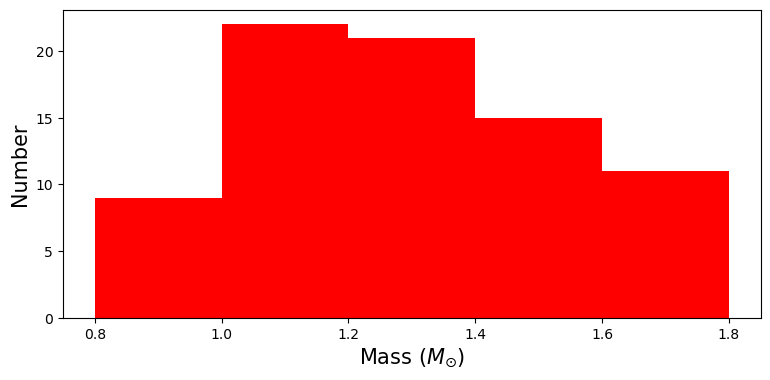}
    \caption{Mass distribution for our sample. As expected, these subgiants tend to be low mass stars ($1 M_{\bigodot}-1.3 M_{\bigodot}$) near the solar age. The values range from 0.75-1.90 $M_{\bigodot}$, with an average uncertainty of 0.21 $M_{\bigodot}$.
    \label{fig:mass}}
\end{figure}

Given their position on the HR diagram, we expected our seismic masses to fall within a range that was typical for stars on the subgiant branch, 0.9-1.6 $M_{\bigodot}$. After our corrections, we can see results that were consistent with these expectations, including a slight skew towards lower masses, likely coming from their longer time in this phase (Figure \ref{fig:mass}), with an uncertainty of 0.21 $M_{\bigodot}$. Seismic mass uncertainties were calculated, as previously mentioned for seismic radius, by propagating forward the uncertainties on the $\nu_{\text{max}}$, $\Delta\nu$, $T_{\text{eff}}$, $f_{\nu_{\text{max}}}$, and $f_{\Delta\nu}$.

\subsection{Major Outlier: TIC 366490487} \label{subsec:outlier}

\begin{figure}[ht!]
\plotone{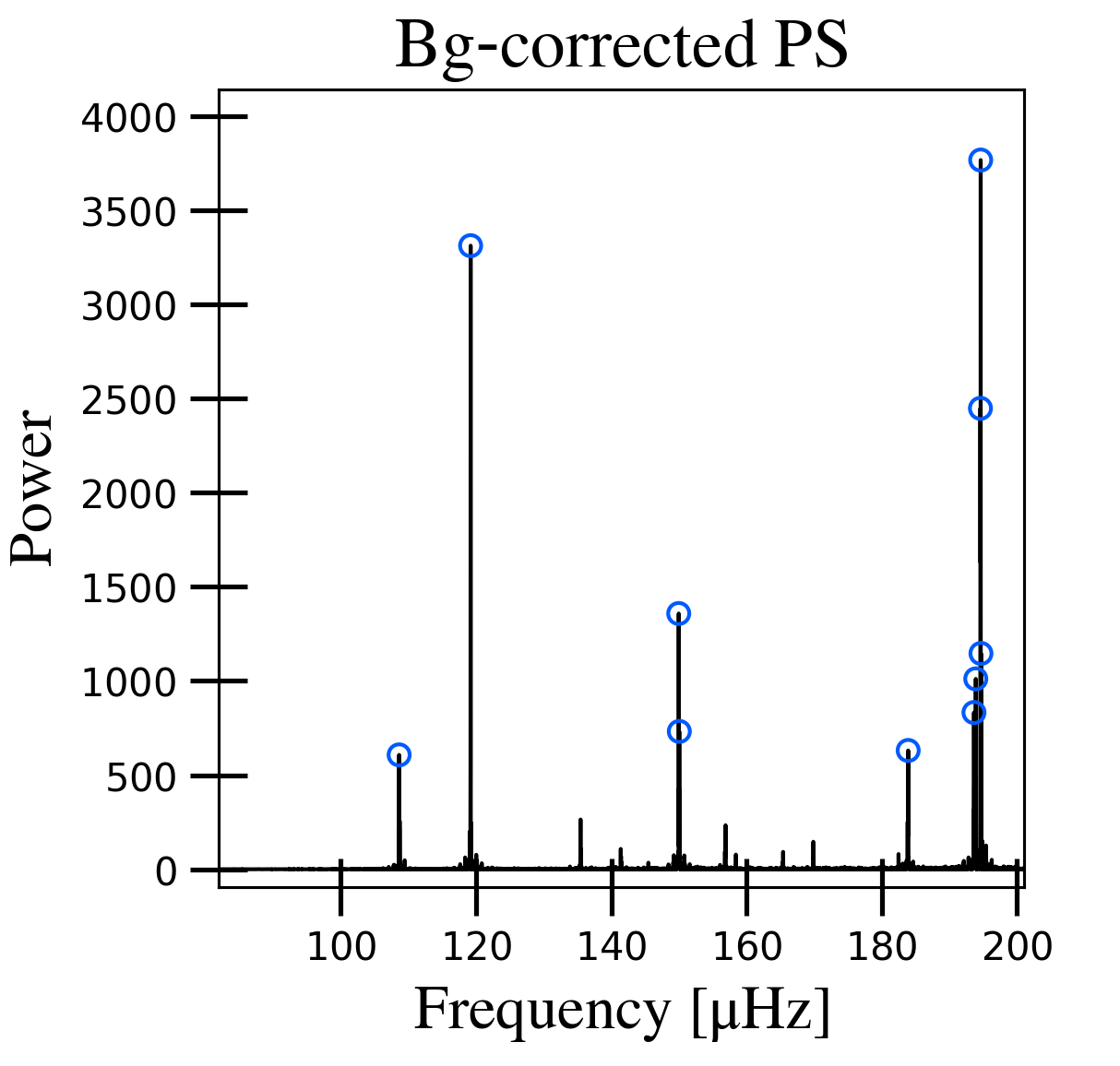}
\caption{The background corrected power spectrum for TIC 366490487. Here we do not have a Gaussian envelope of power but instead we can see multiple large power spikes throughout a large range of frequencies. \label{fig:odd}}
\end{figure}

For our sample, we found one major outlier, TIC 366490487. When completing the initial run through with \texttt{pySYD} star processing, we kept all of the stars that showed an oscillation bump and well-fit $\nu_{\text{max}}$ and $\Delta\nu$ values. One of these stars showed an excess of power not consistent with the expected Gaussian envelope (Figure \ref{fig:odd}). We examined the Cycle 1 and Cycle 3 data for this star separately and found that while the same frequencies were present in each year, the amplitudes and amplitude ratios varied from year to year. Since there was still an excess of power in some capacity, we kept this star in our data table. However, for every validation figure, this particular star would be offset from the rest of the data if we assumed it was a solar-like oscillator. Because of this, we have left this star out of our final figures, but we include it here in case its behavior is interesting or relevant to other research.

\section{Age Comparisons from Interpolating Stellar Evolutionary Tracks}
\label{sec:models}
\subsection{Stellar Model Grid Interpolation} \label{subsec:kiauhoku}

With our seismic masses and surface gravities, photometric effective temperatures, and corrected spectroscopic metallicities, we can utilize the \texttt{Kiauhoku} python package \citep{2020ApJ...888...43C}. 

This package allows previously generated grids of stellar models to be resampled into equivalent evolutionary points \citep{2016ApJS..222....8D} and then interpolates values between the tracks. Unlike some previous works where the models are treated as priors and many stellar parameters have been used as inputs to a grid-based modeling, here we are strictly interpolating within the tracks using our specified input parameters. Specifically we use the \texttt{$StarGridInterpolator.gridsearch_fit$} method to interpolate within models. For each grid, we search models with intial masses between 0.6 and 2.0 $M_{\bigodot}$, EEPs between the zero age main sequence (EEP 201) and the red giant branch bump (EEP 605), and initial metallicities between -1.0 and 0.5. This method uses a Nelder-Mead \citep{10.1093/comjnl/7.4.308} optimizer until a user-specified loss threshold is reached, here we require a threshold of 0.001. Kiauhoku has previously been used for several different grids of stellar evolution models \citep{2022ApJ...927...31T} which can now be loaded automatically. Unfortunately, many of these grids were not computed specifically for asteroseismic work and therefore do not directly track the $\nu_{\text{max}}$ and $\Delta\nu$ of each model point. Instead, we interpolate using our seismically inferred mass and surface gravity as well as our metallicity (\ref{subsec:cal/corr}) in order to infer the expected temperature and age of a star with these input properties. Then, we compare a variety of public model grid's predictions to our measured values. Those comparisons show how well each model matches the data, and at what point they start to differ from these asteroseismic results.

\subsection{Models} \label{subsec:describe}

The models we used to compare to our seismic data were computed with the Modules for Experiments in Stellar Astrophysics (MESA) \citep{2011ApJS..192....3P} Isochrones and Stellar Tracks (MIST) \citep{2016ApJ...823..102C}, the Yale Rotating Stellar Evolution Code (YREC) \citep{1989ApJ...338..424P} with the tracks from \citet{2022ApJ...927...31T}, the Garching Stellar Evolution Code (GARSTEC) \citep{2008Ap&SS.316...99W} with tracks similar to \citet{2013MNRAS.429.3645S}, and the Dartmouth Stellar Evolution Database (DSEP) \citep{2008ApJS..178...89D}.

We chose these models because of their popularity and because they were all previously included in the \texttt{Kiauhoku} package. \citet{2022ApJ...927...31T} gives a brief summary of the input physics used for these models. For MIST, the convective overshoot is diffusive with a value of 0.016 for the core and 0.0174 for the envelope. The solar atmosphere is detailed in \citet{1993sssp.book.....K}. The solar values for X, Y, and Z are 0.7154, 0.2703, and 0.0142, respectively. For DSEP, the model includes convective overshoot with a step of 0.2 $H_p$. The atmosphere is detailed in \citet{1999ApJ...512..377H} and \citet{1999ApJ...525..871H}. The solar values for X, Y, and Z are 0.7071, 0.27402, and 0.01885, respectively. For GARSTEC, the convective overshoot is diffusive with a value of 0.02, the atmosphere is gray, and the solar values for X, Y, and Z, are 0.7090, 0.2716, and 0.0193, respectively. For YREC, the model includes convective overshoot with a step of 0.16 $H_p$, the atmosphere is gray, and the solar values for X, Y, and Z, are 0.709452, 0.2725693, and 0.0179492, respectively. With our parameters from these four models, we are spanning a wide range of options. Although we are not trying every possible option, and we expect that models that are a better match to this specific data could be made in the future, we believe that by using these several different grids we have a representation of how our data compare to models in general, including some estimate of the systematic uncertainties.  
The MIST and DSEP stellar models are both made to cover a large range of stellar properties. The GARSTEC tracks have been often used for seismic analysis, and the YREC stellar model has an emphasis on helio- and asteroseismology and this set of tracks was particularly calibrated to the \citet{2018ApJS..239...32P} giants.

For our interpolation, we use as input parameters: mass, surface gravity, and metallicity, we set our step size to 0.001 $M_{\bigodot}$, 0.001 dex, and 0.001 respectively, expecting the returned output parameters to fit better than this. This is consistent with previous \texttt{Kiauhoku} work, ensuring we were targeting our exact stars.

\subsection{Temperature Comparison} \label{subsec:results}

\begin{figure*}
    \centering
    \includegraphics[width=0.8\textwidth]{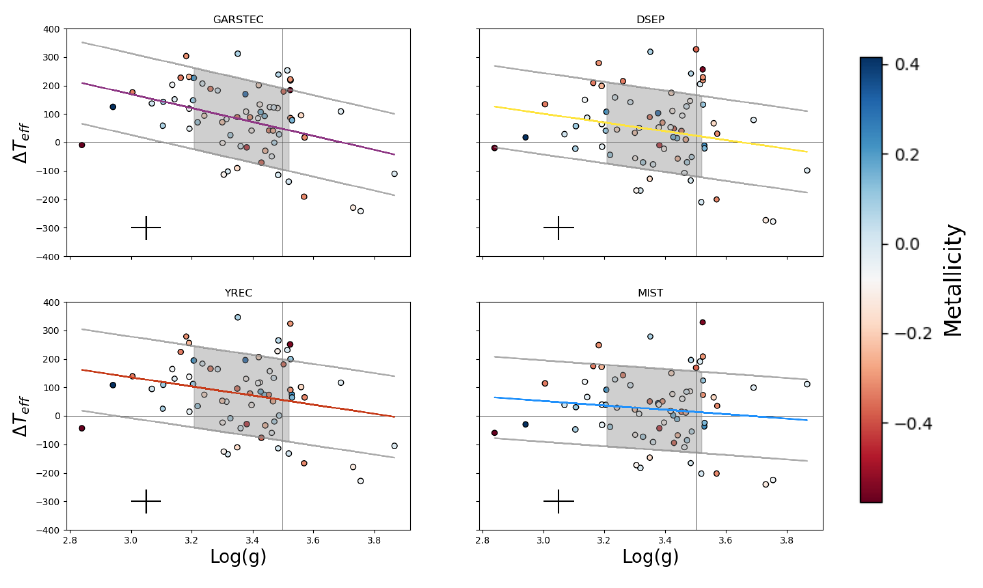}
    \caption{Surface gravity versus the temperature difference between the model and our observables, color coded by metallicity, for each set of models: GARSTEC, DSEP, YREC, and MIST. The first gray line from the linear regression line shows a $1\sigma$ error. The gray box is showing where the bulk of the data is. The horizontal line is at $\Delta T_{\text{eff}}=0$ and the vertical line is at $log(g)=3.5$ roughly separating the giants on the left and the subgiants on the right. Our observed errors are shown in the bottom left corner of each subplot.}
    \label{fig:model comp}
\end{figure*}

We took the difference in our effective temperature and the model's effective temperature and plotted that against surface gravity. In Figure \ref{fig:model comp}, we look at $\Delta T_{\text{eff}}$ of -500 to 500 K to focus on the best-fit model. On average, the models differed by about $\pm$ 125 K from our data. All of the models gave fairly consistent results for the temperature offsets. Similar results were obtained for the ages (see \ref{subsec:ages}).

The MIST model had the lowest average temperature difference (i.e., photometric $T_{\text{eff}}$ - model $T_{\text{eff}}$) with our data at $\pm$ 98 K, while the other YREC, DSEP, and GARSTEC had an average difference of about $\pm$ 124 K, $\pm$ 112 K, and $\pm$ 123 K respectively. The standard deviation from the linear regression line for MIST, YREC, DSEP, and GARSTEC was 94 K, 92 K, 95 K, and 93 K respectively and the average difference in temperature between the models was relatively low ($\pm$ 79 K).

At log(g) between 2.8 and 3.6, we found that the difference in $T_{\text{eff}}$ between the data and the models was fairly low, with an average of $\pm$ 114K from our data. However, when including log(g) values greater than 3.6, the average difference in $T_{\text{eff}}$ rose to about $\pm$ 230 K.

These offsets show that the stellar model grids are roughly consistent with, but not exactly calibrated to, giant stars, with a trend of predicting that they will be hotter than what we observe. For the true subgiant stars, the models are not very accurate, with the average difference being double of what it was for the lower giants. The scatter between the points also increases, meaning it would be much more challenging to calibrate subgiant stellar models to photometric and spectroscopic data we have already. More generally, we expect that stellar models need to be improved for all evolutionary stages, and especially for transition stages like the subgiant phase where the changes in the internal structure of a star are much more pronounced.

\subsection{Stellar Ages} \label{subsec:ages}

\begin{figure}[ht!]
\includegraphics[width=0.45\textwidth,clip=true]{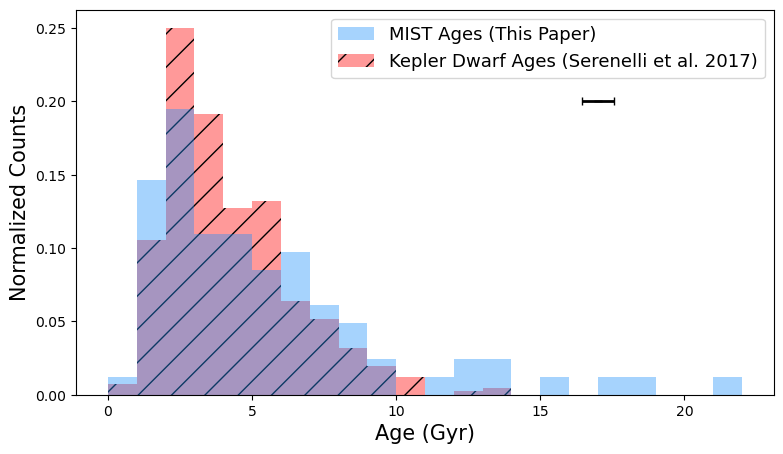}
    \caption{Age estimates from the MIST stellar model tracks. We show our ages (blue) normalized in comparison to the ages from \citet{2017ApJS..233...23S} (red) and can see that the ages for the \tess\ stars are quite consistent with previous studies from {\it Kepler}, perhaps because both sample a local population. The average observed error bar is in the upper right corner.}
    \label{fig:ages}
\end{figure}

Alongside our ability to compare our surface gravities and effective temperatures with those of the models, we were also able to compute age estimates for all of our stars, using each set of models. We found that the age estimates for each star were relatively similar between all four of the models so we chose to only display the results from the MIST model for clarity (Figure \ref{fig:ages}). A comprehensive list of all the ages from the 4 models is given in Table \ref{tab:table2}.

The ages range from about 1.2-21.2 Gyr, with the bulk of our stars from 1.2-9.6 Gyr with an average observational uncertainty of 0.72 Gyr. When determining our final ages and uncertainties, we found that there needed to be errors to account for the difference in ages given by the four models, as well as the errors we have on our mass and metallicity. For completeness, we have added these intermediate uncertainties and the final uncertainty for each star. $\sigma_{\text{Age}_{\text{{the}}}}$ is the average theoretical error given by \texttt{Kiauhoku} between the 4 model grids. $\sigma_{\text{Age}_{\text{{obs}}}}$ is the average observational error. For each individual star this was calculated by running Kiauhoku with $M_{\text{cor}}$ + 1$\sigma_{\text{M}_{\text{cor}}}$ and  [M/H] - 1$\sigma_{\text{[M/H]}}$ to get the lower age limit, and running Kiauhoku with $M_{\text{cor}}$ - 1$\sigma_{\text{M}_{\text{cor}}}$ and  [M/H] + 1$\sigma_{\text{[M/H]}}$ to get the higher age limit, then averaging these two values. The lower and higher age limits are taken from the mean age of the four models given by Kiauhoku. The total uncertainty, $\sigma_{\text{Age}}$, is the average theoretical and observational errors of the four stellar model grids added in quadrature. We list the total uncertainty in Table \ref{tab:table1} and the separate theoretical and observational uncertainties in Table \ref{tab:table2}.

In our histogram in Figure \ref{fig:ages}, we show our ages compared to the ages given by \citet{2017ApJS..233...23S}. We observe that the ages of our stars predicted by the models align with those of nearby stars observed by {\it Kepler}, affirming that our asteroseismic findings allow us to estimate ages that are consistent with earlier studies. Thus, this asteroseismic work, as well as future works, can be combined and used to further our understanding of galactic formation and evolution.

\subsubsection{MIST Age Comparisons} \label{subsubsec:age}

\begin{figure*}[t!]
    \centering
    \begin{subfigure}[]
        \centering
    \includegraphics[width=0.45\textwidth,clip=true, trim=0.4in 0in 1.0in 0.5in]{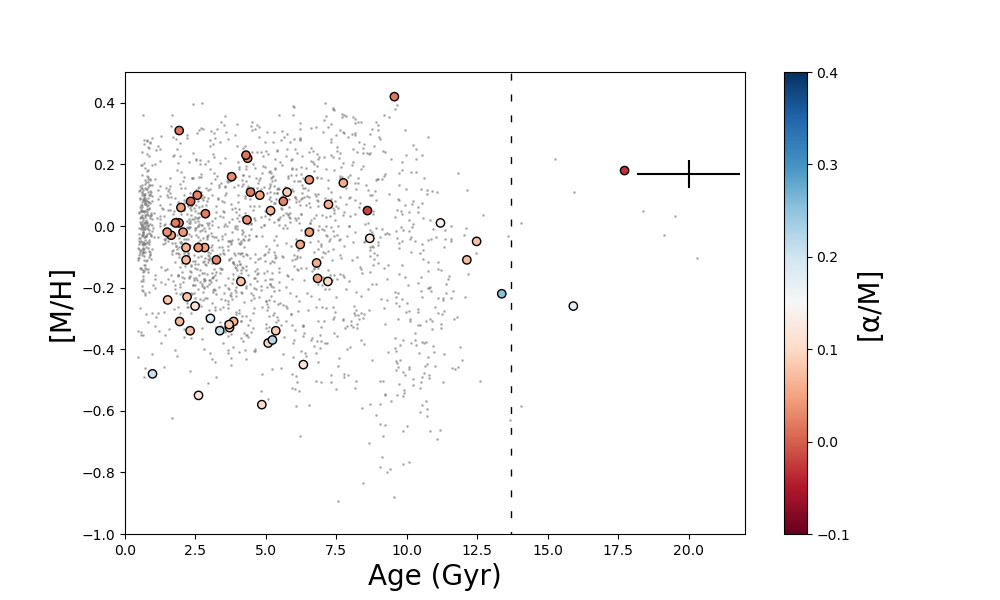}
    \label{subfig:met_age}
    \end{subfigure}
    ~ 
    \begin{subfigure}[]
        \centering
    \includegraphics[width=0.45\textwidth,clip=true, trim=0.4in 0in 1.0in 0.5in]{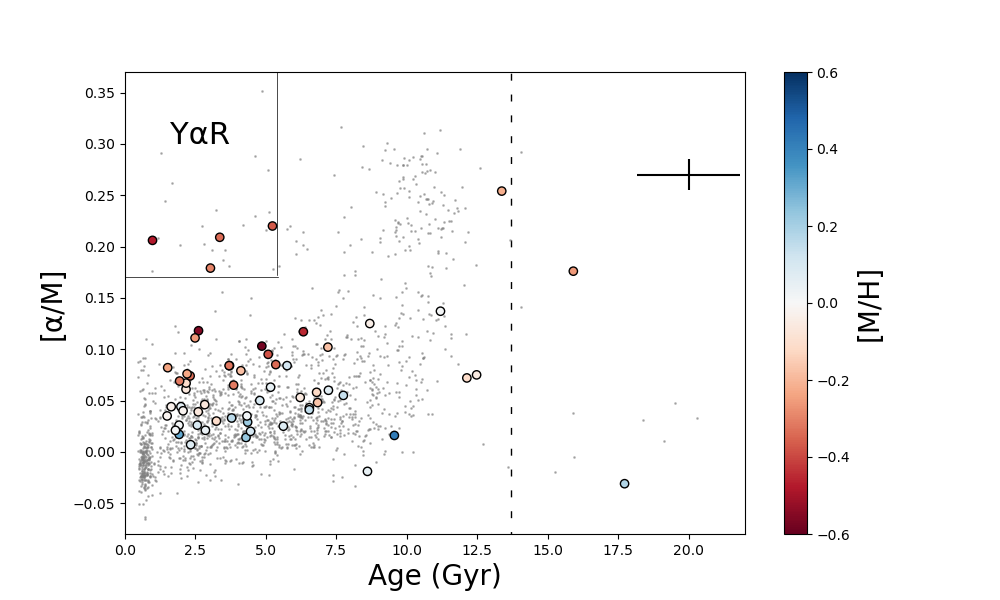}
    \label{subfig:alpha_age}
    \end{subfigure}
    \caption{We show for comparison (gray) the full sample from the APOGEE DR17 spectroscopic survey in this temperature and metallicity regime. {(LEFT)}: Our data is circled in black and color-coded by [$\alpha$/M]. [M/H] plotted against ages given by the MIST stellar models. We see no strong correlation between [M/H] and age.
    \textbf{(RIGHT)}: Our data is circled in black and color-coded by [M/H]. [$\alpha$/M] plotted against ages given by the MIST stellar model. In the top left corner, we can see our young alpha-enhanced (y$\alpha$r) stars. The $\alpha$-rich, thick disk stars are in the top right quadrant, with the remaining lower half of the chart representing the thin disk. Most of the sample are low-alpha stars. The dotted line shows the age of the universe for reference. Average error bars are shown in the top right corner.}
    \label{fig:metandalpha_age}
\end{figure*}

\begin{figure}[ht!]
\includegraphics[width=0.47\textwidth,clip=true, trim=0.4in 0in 1.0in 0.5in]{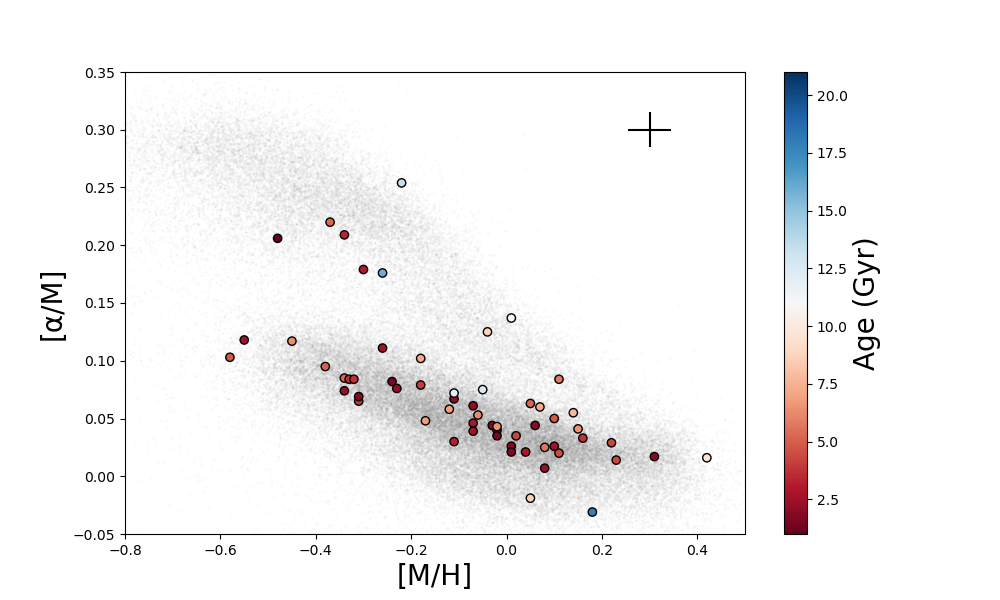}
    \caption{[$\alpha$/M] plotted against [M/H]. Our data is circled in black and color-coded by age. We show for comparison (gray) the full sample from the APOGEE DR17 spectroscopic survey in this temperature and metallicity regime. Average error bars are shown in the top right corner.}
    \label{fig:contour_age}
\end{figure}

We used our metallicities and the alpha-enhancement parameter, [$\alpha$/M], to look for trends between age and composition. For some of our stars, APOGEE did not have published data. We kept the stars with no [$\alpha$/M] value out of Figures \ref{fig:metandalpha_age} and \ref{fig:contour_age} and listed an [$\alpha$/M] value of 0 in Table \ref{tab:table1}. We show the results for MIST in Figures \ref{fig:metandalpha_age} and \ref{fig:contour_age}.

In Figure \ref{fig:metandalpha_age}a, we show the APOGEE metallicities versus the ages obtained using MIST. We do not see strong trends between [M/H] and age in this regime, reinforcing that [M/H] cannot be used as an age indicator for metal-rich populations.

Most stars are between the ages of 1.2 and 9.6 Gyr, which is as expected for a thin disk, alpha-poor population, with the median age of 4.4 Gyr. However, 6 of our 82 stars turned out much older than the typical range for subgiant stars. In Figure \ref{fig:metandalpha_age}a, it is shown that the metallicities for these 6 stars are typical, ranging from about -0.3 to 0.2, and they were well fit by our modeling. This is not the first time something like this has been seen. For example, \citet{2009IAUS..258..161S} shows how eclipsing binaries give very accurate physical property values, but the ages can show these stars to be 50-90\% older than their actual age. This happens when a close binary system strips mass from its counterpart, leading to a less massive subgiant than what a star that age would normally be, so we suggest that these stars may have undergone mass loss because of an interaction with a binary companion. However, we caution that our age uncertainties in this regime are significant, with only two stars falling $1\sigma$ above the age of the universe. Therefore, we cannot exclude that other factors, such as modeling uncertainties or observational errors, could be a factor. In Figure \ref{fig:metandalpha_age}b, we show no strong correlation between the ages of our stars and their $\alpha$-element enhancement. We see the same 6 outliers. We also identify 4 stars that would be classified as young $\alpha$ rich stars by \citet{2015MNRAS.451.2230M} (top-left box in Figure \ref{fig:metandalpha_age}b). It has been suggested that these could either result from recent formation of stars from $\alpha$-enriched gas or that they could be the result of mass gain from close interaction with a binary companion, making them appear younger than they actually are \citep{2015A&A...576L..12C, 2023A&A...671A..21J}. We expect that the existence and frequency of such stars before the giant branch could be helpful in illuminating the dominant formation mechanism.

Figure \ref{fig:contour_age} shows [$\alpha$/M] plotted against [M/H] color-coded by age. The gray points in the background are stars from APOGEE DR17 in the same metallicity and temperature regime as our sample. With this underlay of data, we can see that most of our stars fall on the alpha-poor (thin-disk) band, although we do have a few stars lying on the alpha-rich (thick-disk) band. Comparing our ages to those expected from galactic chemical enrichment models, we notice both young $\alpha$ rich stars as well as at least one star that is very old and $\alpha$ poor. The number of age outliers in key places makes it challenging to place strong chemical enrichment constraints from just this small sample, but we suggest that larger samples of stars like this could be helpful in understanding the chemical evolution of the Milky Way.

\section{Discussions and Conclusions} \label{sec:con}

In this paper, we present a detailed first look at the oscillating subgiants and lower giants from the \tess\ Northern and Southern CVZs. We show that two years of data is sufficient to measure the precise $\nu_{\text{max}}$ and $\Delta\nu$ parameters for these stars, and compute seismic radii, masses, surface gravities, and ages. We present seismic results for 82 stars, with masses ranging from 0.75-1.90 $M_{\bigodot}$ with average uncertainties of 0.21 $M_{\bigodot}$, radii ranging from 1.47-6.86 $R_{\bigodot}$ with an uncertainty of 0.27 $R_{\bigodot}$, and surface gravities ranging from 2.82-4.02 dex with an uncertainty of $\approx$0.06 dex. These results presented here are shown to be consistent with previous results, including those from other seismic studies, spectroscopy, and {\it Gaia}.

We demonstrated in Section \ref{sec:models} how our data compares to currently published stellar models and give a first test at how these models do against real data that has not been used before. We compare them to several public model grids in this regime. In general, we found that the models provided predictions that are similar to each other and similar to the data, although we note some slight offsets and trends with surface gravity. We also showed that this sample of stars probes a range of ages, compositions, and galactic populations including young alpha-enhanced, alpha-rich, and alpha-poor disks.

We expect this sample to be useful for a variety of future investigations including the calibration of spectroscopic surface gravity values \citep{2013A&A...556A..59H, 2015ApJ...807....4L}, identifying stars for He analysis \citep{2024ApJ...965..171L}, checking stellar models in this region \citep{2017ApJ...842....1G, 2022AJ....164...14S, 2020MNRAS.495.3431L}, estimating ages and age distributions \citep{2022Natur.603..599X}, identifying mixed modes \citep{2011A&A...535A..91D}, and analyzing core rotation \citep{2012ApJ...756...19D, 2012A&A...548A..10M}. More generally, using \tess\ data for asteroseismology is an effective way to get precise and accurate stellar parameters. As more data is gathered on stellar oscillations, we will have a greater ability to understand structures and compositions of stars along with interesting and important details like convective mixing and magnetic activity within stellar interiors.

\section*{Acknowledgements} \label{sec:acknowledgment}
We thank Ashley Chontos for helpful discussions about using the \texttt{pySYD} package.
We thank Zachary Claytor for help with HiperGator.
We thank Sarbani Basu for helpful suggestions.
We thank Christopher Lindsay and Joel Ong for sending their $\nu_{\text{max}}$ and $\Delta\nu$ values for comparison in this paper.
We thank Dennis Stello for helpful suggestions.

SG and JT acknowledge support from NASA grants 80NSSC23K0143 and 80NSSC23K0436. JT also acknowledges support from NASA grants 80NSSC20K0056 and 80NSSC19K0367. DGR acknowledges support from the Juan de la Cierva program under contract JDC2022-049054-I.

This work has made use of data from the European Space Agency (ESA) mission
{\it Gaia} (\url{https://www.cosmos.esa.int/gaia}), processed by the {\it Gaia}
Data Processing and Analysis Consortium (DPAC,
\url{https://www.cosmos.esa.int/web/gaia/dpac/consortium}). Funding for the DPAC
has been provided by national institutions, in particular the institutions
participating in the {\it Gaia} Multilateral Agreement.

Funding for the Sloan Digital Sky Survey IV has been provided by the Alfred P. Sloan Foundation, the U.S. Department of Energy Office of Science, and the Participating Institutions.

SDSS-IV acknowledges support and resources from the Center for High Performance Computing at the University of Utah. The SDSS website is www.sdss4.org.

SDSS-IV is managed by the Astrophysical Research Consortium for the Participating Institutions of the SDSS Collaboration including the Brazilian Participation Group, the Carnegie Institution for Science, Carnegie Mellon University, Center for Astrophysics | Harvard \& Smithsonian, the Chilean Participation Group, the French Participation Group, Instituto de Astrof\'isica de Canarias, The Johns Hopkins University, Kavli Institute for the Physics and Mathematics of the Universe (IPMU) / University of Tokyo, the Korean Participation Group, Lawrence Berkeley National Laboratory, Leibniz Institut f\"ur Astrophysik Potsdam (AIP),  Max-Planck-Institut f\"ur Astronomie (MPIA Heidelberg), Max-Planck-Institut f\"ur Astrophysik (MPA Garching), Max-Planck-Institut f\"ur Extraterrestrische Physik (MPE), National Astronomical Observatories of China, New Mexico State University, New York University, University of Notre Dame, Observat\'ario Nacional / MCTI, The Ohio State University, Pennsylvania State University, Shanghai Astronomical Observatory, United Kingdom Participation Group, Universidad Nacional Aut\'onoma de M\'exico, University of Arizona, University of Colorado Boulder, University of Oxford, University of Portsmouth, University of Utah, University of Virginia, University of Washington, University of Wisconsin, Vanderbilt University, and Yale University.

Transiting Exoplanet Survey Satellite (TESS) was accessed on January 2023 from https://registry.opendata.aws/tess. This paper includes data collected with the \tess\ mission, obtained from the AWS Open Data copy of the MAST Archive at the Space Telescope Science Institute (STScI) \citep{https://doi.org/10.17909/fwdt-2x66}. Funding for the \tess\ mission is provided by the NASA Explorer Program. STScI is operated by the Association of Universities for Research in Astronomy, Inc., under NASA contract NAS 5–26555.

\software{\texttt{Kiauhoku} \textbf{\citep{2020ApJ...888...43C}}, \texttt{NumPy} \textbf{\citep{harris2020array}}, \texttt{Lightkurve} \textbf{\citep{2018ascl.soft12013L}}, \texttt{AstroPy} \textbf{\citep{astropy:2013, astropy:2018, astropy:2022}}, \texttt{Matplotlib} \textbf{\citep{4160265}}, \texttt{SciPy} \textbf{\citep{2020NatMe..17..261V}}, \texttt{pySYD} \textbf{\citep{2022JOSS....7.3331C}}, \texttt{HeyEchelle} \textbf{\citep{daniel_hey_2020_3629933}}, \texttt{Pandas} \textbf{\citep{mckinney-proc-scipy-2010}}, \texttt{ASFGrid} \textbf{\citep{2016ascl.soft03009S}}}

\bibliographystyle{aasjournal} 
\bibliography{introduction, method, validation, models, conclusion, software}{}

\onecolumngrid
\newpage
\begin{longrotatetable}
\begin{deluxetable}{cccccccccccccccccccc}

\tabletypesize{\footnotesize}
\tablewidth{0pt}

 \tablecaption{Seismic Values and Uncertainties \label{tab:table1}}

 \tablehead{
 \colhead{TIC ID} & \colhead{$\nu_{\text{max}}$} & \colhead{$\Delta\nu$} & \colhead{$f_{\Delta\nu}$} & \colhead{$\log(g)$} & \colhead{$\text{M}_{\text{cor}}$} & \colhead{$\text{R}_{\text{cor}}$} & \colhead{Age} & \colhead{$\text{T}_{\text{eff}}$} & \colhead{[M/H]} & \colhead{[$\alpha/\text{M}$]} & \colhead{$\sigma_{\nu_{\text{max}}}$} & \colhead{$\sigma_{\Delta\nu}$} & \colhead{$\sigma_{\log(g)}$} & \colhead{$\sigma_{\text{M}_{\text{cor}}}$} & \colhead{$\sigma_{\text{R}_{\text{cor}}}$} & \colhead{$\sigma_{\text{Age}}$} & \colhead{$\sigma_{\text{T}_{\text{eff}}}$} & \colhead{$\sigma_{\text{[M/H]}}$} & \colhead{$\sigma_{[\alpha/\text{M}]}$}}
 

\startdata
25156036	&	213.91	&	15.64	&	0.985	&	3.21	&	1.37	&	4.72	&	4.37	&	4652	&	0.22	&	0.029	&	6.9	&	0.29	&	0.061	&	0.14	&	0.3	&	0.26	&	24	&	0.0057	&	0.0045	\\
31506424	&	421.87	&	26.49	&	1.015	&	3.52	&	1.34	&	3.24	&	2.79	&	5217	&	-0.55	&	0.118	&	8.06	&	2.21	&	0.015	&	0.08	&	0.32	&	0.13	&	26	&	0.0096	&	0.11	\\
38602419	&	428.03	&	29.11	&	1.016	&	3.57	&	0.97	&	2.72	&	10.1	&	5256	&	-0.32	&	0	&	9.46	&	0.83	&	0.036	&	0.2	&	0.08	&	4.42	&	43.5	&	0.0158	&	0	\\
41587424	&	194.37	&	14.18	&	0.999	&	3.19	&	1.56	&	5.22	&	2.33	&	4914	&	-0.07	&	0.061	&	4.71	&	0.76	&	0.076	&	0.15	&	0.45	&	0.21	&	29	&	0.0069	&	0.0056	\\
55270123	&	406.58	&	24.86	&	1.006	&	3.48	&	1.54	&	3.55	&	2.36	&	5067	&	-0.11	&	0.067	&	17.36	&	1.41	&	0.061	&	0.24	&	0.42	&	0.51	&	23	&	0.0074	&	0.0059	\\
141335100	&	96.69	&	8.97	&	0.984	&	2.94	&	1.12	&	6.41	&	9.34	&	4508	&	0.42	&	0.016	&	15.97	&	0.64	&	0.02	&	0.15	&	0.39	&	1.59	&	16	&	0.005	&	0.0038	\\
141626634	&	491.57	&	28.16	&	1.017	&	3.65	&	1.7	&	3.36	&	1.79	&	5309	&	0	&	0	&	7	&	0.14	&	0.037	&	0.18	&	0.03	&	0.16	&	34.5	&	0.01	&	0	\\
141757732	&	136.77	&	11.1	&	0.994	&	3.16	&	1.46	&	6.03	&	2.49	&	4885	&	-0.34	&	0.074	&	18.06	&	1.23	&	0.073	&	0.14	&	0.65	&	0.17	&	16	&	0.0082	&	0.0069	\\
142109390	&	382.05	&	22.98	&	1.008	&	3.48	&	1.65	&	3.83	&	2.1	&	4892	&	0.06	&	0.044	&	2.23	&	2.5	&	0.058	&	0.44	&	0.52	&	3.84	&	21.5	&	0.0064	&	0.0051	\\
149390710	&	329.53	&	23.43	&	1.01	&	3.44	&	1.02	&	3.21	&	9.11	&	5046	&	-0.24	&	0	&	2.37	&	0.37	&	0.047	&	0.05	&	0.07	&	0.6	&	26.5	&	0.0396	&	0	\\
150030411	&	432.82	&	26.86	&	0.994	&	3.53	&	1.36	&	3.26	&	4.37	&	4895	&	0.23	&	0.014	&	6.03	&	1.31	&	0.059	&	0.14	&	0.3	&	0.38	&	22.5	&	0.0057	&	0.0406	\\
150062447	&	650.52	&	41.12	&	1.015	&	3.73	&	0.83	&	2.05	&	15.78	&	5124	&	-0.13	&	0	&	17.23	&	1.39	&	0.045	&	0.26	&	0.29	&	16.05	&	34	&	0.0013	&	0	\\
150166759	&	210.19	&	15.58	&	1.005	&	3.22	&	1.29	&	4.58	&	4.93	&	4836	&	0.1	&	0.05	&	1.76	&	0.47	&	0.069	&	0.17	&	0.3	&	0.78	&	21.5	&	0.006	&	0.0049	\\
150393198	&	188.52	&	14.94	&	1.015	&	3.19	&	1.06	&	4.38	&	9.36	&	4835	&	-0.04	&	0.125	&	6.7	&	0.38	&	0.029	&	0.1	&	0.04	&	0.81	&	17	&	0.0064	&	0.0053	\\
150430044	&	310.43	&	19.48	&	0.999	&	3.38	&	1.75	&	4.38	&	1.92	&	4852	&	0.31	&	0.017	&	10.37	&	0.43	&	0.039	&	0.48	&	0.36	&	1.81	&	17.5	&	0.0056	&	0.0043	\\
150442152	&	231.97	&	17.27	&	0.995	&	3.33	&	1.22	&	4.23	&	5.84	&	4902	&	0.08	&	0.025	&	33.14	&	1.09	&	0.074	&	0.2	&	0.04	&	1.58	&	26	&	0.0064	&	0.0129	\\
167342488	&	345.87	&	22.38	&	1.001	&	3.46	&	1.44	&	3.74	&	3.07	&	4982	&	-0.07	&	0.046	&	4.98	&	0.69	&	0.062	&	0.05	&	0.35	&	0.17	&	19.5	&	0.0072	&	0.0057	\\
167548586	&	412.55	&	26.5	&	0.999	&	3.49	&	1.24	&	3.18	&	5.42	&	4959	&	0.05	&	0.063	&	28.06	&	0.51	&	0.064	&	0.16	&	0.03	&	1.03	&	28.5	&	0.0065	&	0.0052	\\
177241155	&	200.91	&	14.58	&	1.007	&	3.19	&	1.47	&	4.99	&	2.61	&	4855	&	-0.25	&	0	&	6.09	&	1.48	&	0.073	&	0.06	&	0.49	&	0.15	&	18.5	&	0.0127	&	0	\\
220457103	&	307.65	&	21.81	&	1.014	&	3.43	&	1.13	&	3.49	&	5.5	&	5201	&	-0.38	&	0.095	&	11.84	&	0.63	&	0.031	&	0.18	&	0.02	&	1.48	&	17.5	&	0.0091	&	0.12	\\
257720189	&	362.35	&	24.4	&	1.005	&	3.47	&	1.13	&	3.25	&	7.54	&	4938	&	0.07	&	0.06	&	183.11	&	11.92	&	0.058	&	0.15	&	0.05	&	1.2	&	16.5	&	0.0063	&	0.0469	\\
260078030	&	560.81	&	32.72	&	1.02	&	3.65	&	1.42	&	2.85	&	2.74	&	5436	&	-0.26	&	0.111	&	32.78	&	0.18	&	0.023	&	0.22	&	0.35	&	0.67	&	53.5	&	0.0444	&	0.0623	\\
260655847	&	809.66	&	47.24	&	1.02	&	3.87	&	1.13	&	2.07	&	6.98	&	5506	&	-0.01	&	0	&	7.19	&	0.43	&	0.018	&	0.06	&	0.09	&	0.34	&	54	&	0.0192	&	0	\\
271977529	&	444.19	&	27.47	&	0.989	&	3.52	&	1.32	&	3.2	&	4.59	&	4790	&	0.11	&	0.02	&	20.92	&	0.93	&	0.049	&	0.17	&	0.19	&	0.44	&	20.5	&	0.0061	&	0.0048	\\
278731442	&	273.31	&	20.35	&	1.001	&	3.39	&	1.02	&	3.56	&	8.72	&	4947	&	-0.3	&	0	&	36.2	&	0.56	&	0.045	&	0.14	&	0.13	&	1.62	&	20	&	0.0124	&	0	\\
279055252	&	401.04	&	24.04	&	1.009	&	3.47	&	1.64	&	3.71	&	2.04	&	5012	&	0.01	&	0.026	&	30.97	&	0.4	&	0.007	&	0.35	&	0.23	&	0.82	&	27	&	0.0056	&	0.0053	\\
279090824	&	565.79	&	37.84	&	1.018	&	3.68	&	1.48	&	2.91	&	2.43	&	5371	&	-0.23	&	0.076	&	11.93	&	0.31	&	0.021	&	0.11	&	0.18	&	0.18	&	30	&	0.0093	&	0.0285	\\
279571746	&	234.48	&	16.06	&	0.994	&	3.29	&	1.68	&	4.94	&	1.9	&	4882	&	0.01	&	0.021	&	13.08	&	0.16	&	0.023	&	0.39	&	0.34	&	1.77	&	17.5	&	0.0067	&	0.0053	\\
293271686	&	275.54	&	18.8	&	1.005	&	3.36	&	1.46	&	4.21	&	2.96	&	5051	&	0	&	0	&	5.81	&	0.98	&	0.068	&	0.12	&	0.3	&	0.26	&	28	&	0.01	&	0	\\
300085625	&	330.05	&	22.73	&	0.996	&	3.44	&	1.13	&	3.43	&	7.94	&	4811	&	0.14	&	0.055	&	4.4	&	0.21	&	0.047	&	0.16	&	0.13	&	1.35	&	20	&	0.0059	&	0.0046	\\
300088321	&	1201.46	&	67.28	&	1.02	&	4.01	&	0.82	&	1.47	&	14.21	&	5614	&	-0.03	&	0	&	23.69	&	2.54	&	0.04	&	0.23	&	0.19	&	12.17	&	47.5	&	0.0312	&	0	\\
300161002	&	409.07	&	25.5	&	0.991	&	3.51	&	1.39	&	3.41	&	3.64	&	4816	&	0	&	0	&	8.12	&	0.27	&	0.039	&	0.22	&	0.19	&	0.67	&	20	&	0.01	&	0	\\
300444376	&	483.47	&	30.39	&	1.01	&	3.61	&	1.71	&	3.21	&	1.74	&	5135	&	-0.03	&	0.044	&	4.93	&	1.9	&	0.073	&	0.38	&	0.1	&	1.23	&	33	&	0.0074	&	0.0056	\\
349059821	&	404.29	&	25.59	&	1.002	&	3.52	&	1.35	&	3.34	&	3.3	&	5017	&	-0.3	&	0.179	&	5.9	&	2.02	&	0.031	&	0.1	&	0.48	&	0.19	&	28	&	0.0077	&	0.0696	\\
349374407	&	224.93	&	17.03	&	1.001	&	3.26	&	1.14	&	4.15	&	5.83	&	4874	&	-0.34	&	0.085	&	1.04	&	0.17	&	0.069	&	0.19	&	0.07	&	1.42	&	16.5	&	0.0083	&	0.0069	\\
349374677	&	185.34	&	14.37	&	0.99	&	3.18	&	1.26	&	4.85	&	4.17	&	4757	&	-0.31	&	0.065	&	6.39	&	0.71	&	0.061	&	0.09	&	0.27	&	0.25	&	13.5	&	0.0079	&	0.0067	\\
349477633	&	287.16	&	19.68	&	0.999	&	3.38	&	1.37	&	4.02	&	3.51	&	4963	&	-0.11	&	0.03	&	1.18	&	0.6	&	0.069	&	0.16	&	0.06	&	0.35	&	17	&	0.0073	&	0.0058	\\
349973705	&	366.58	&	23.06	&	1.015	&	3.48	&	1.54	&	3.71	&	2.47	&	5219	&	0	&	0	&	6.51	&	1.47	&	0.062	&	0.17	&	0.19	&	0.35	&	47.5	&	0.01	&	0	\\
350297922	&	286.97	&	20.99	&	1.013	&	3.38	&	1.06	&	3.49	&	6.79	&	5136	&	-0.45	&	0.117	&	4.53	&	0.62	&	0.075	&	0.11	&	0.09	&	0.76	&	37.5	&	0.0089	&	0.11	\\
350335258	&	113.69	&	9.87	&	0.998	&	3.0054	&	1.31	&	6.27	&	3.61	&	4844	&	-0.34	&	0.209	&	6.95	&	0.31	&	0.075	&	0.21	&	0.42	&	0.86	&	21	&	0.0067	&	0.0066	\\
350343922	&	258.53	&	19.28	&	1.008	&	3.35	&	1.09	&	3.75	&	7.37	&	5085	&	-0.17	&	0.048	&	11.32	&	0.11	&	0.024	&	0.02	&	0.05	&	0.42	&	29	&	0.0074	&	0.0061	\\
350443250	&	456.92	&	27.22	&	1.02	&	3.56	&	1.52	&	3.3	&	2.12	&	5263	&	-0.31	&	0.069	&	18.93	&	1.46	&	0.042	&	0.09	&	0.25	&	0.14	&	32	&	0.0062	&	0.12	\\
350585437	&	247.88	&	18.55	&	1.009	&	3.32	&	1.09	&	3.85	&	8.07	&	5015	&	0	&	0	&	6.05	&	0.34	&	0.06	&	0.1	&	0.15	&	1.04	&	27.5	&	0.01	&	0	\\
374858999	&	486.23	&	28.04	&	1.012	&	3.58	&	1.65	&	3.34	&	1.91	&	5194	&	-0.07	&	0	&	15.43	&	0.91	&	0.054	&	0.34	&	0.41	&	0.67	&	29	&	0.0104	&	0	\\
375089600	&	350.67	&	24.85	&	1.008	&	3.46	&	0.95	&	3.02	&	13.49	&	4947	&	-0.05	&	0.075	&	5.16	&	0.34	&	0.031	&	0.16	&	0.13	&	3.08	&	17.5	&	0.0067	&	0.0055	\\
381976956	&	79.97	&	7.88	&	1.011	&	2.84	&	1.12	&	6.86	&	5.17	&	5022	&	-0.58	&	0.103	&	0.7	&	0.21	&	0.099	&	0.27	&	0.14	&	2.89	&	22.5	&	0.009	&	0.0081	\\
382067256	&	284.49	&	18.02	&	0.991	&	3.35	&	1.9	&	4.77	&	1.36	&	4842	&	0.08	&	0	&	9.31	&	1.32	&	0.051	&	0.7	&	0.63	&	0.91	&	31.5	&	0.0125	&	0	\\
382159212	&	536.54	&	35.3	&	1.014	&	3.65	&	1.55	&	3.08	&	2.42	&	5233	&	0	&	0	&	25.99	&	0.22	&	0.025	&	0.29	&	0.43	&	0.73	&	25.5	&	0.01	&	0	\\
382302322	&	404.23	&	26.67	&	1.005	&	3.52	&	1.14	&	3.07	&	5.74	&	5051	&	-0.31	&	0	&	9.51	&	1.67	&	0.063	&	0.16	&	0.06	&	1.05	&	26	&	0.0021	&	0	\\
198383789	&	342.47	&	23.33	&	1.007	&	3.42	&	1.11	&	3.33	&	7.33	&	4867	&	-0.12	&	0.058	&	24.96	&	0.18	&	0.059	&	0.16	&	0.04	&	1.23	&	18	&	0.0071	&	0.0059	\\
219096868	&	178.07	&	13.18	&	0.994	&	3.14	&	1.61	&	5.55	&	2.19	&	4848	&	-0.02	&	0.04	&	1.55	&	0.06	&	0.077	&	0.27	&	0.68	&	0.57	&	16.5	&	0.0058	&	0.0053	\\
219792330	&	433.4	&	28	&	1.016	&	3.56	&	1.08	&	2.9	&	7.85	&	4954	&	-0.18	&	0.102	&	15.55	&	1.01	&	0.017	&	0.12	&	0.08	&	0.8	&	15	&	0.0061	&	0.0946	\\
219801618	&	273.68	&	17.6	&	1.013	&	3.35	&	1.88	&	4.76	&	1.02	&	5186	&	-0.48	&	0.206	&	15.55	&	1.01	&	0.064	&	0.42	&	0.57	&	0.9	&	25.5	&	0.0109	&	0.19	\\
219856298	&	444.96	&	28.67	&	1	&	3.5	&	1.14	&	2.93	&	5.67	&	4976	&	-0.37	&	0.22	&	15.55	&	1.01	&	0.058	&	0.15	&	0.05	&	0.89	&	18	&	0.008	&	0.0667	\\
224601836	&	620.62	&	37.88	&	1.017	&	3.76	&	1.04	&	2.34	&	9.18	&	5308	&	-0.06	&	0	&	6.86	&	0.84	&	0.025	&	0.17	&	0.13	&	2.7	&	28	&	0.0051	&	0	\\
224605278	&	243.03	&	18.08	&	0.996	&	3.31	&	1.16	&	4.02	&	6.68	&	4876	&	-0.06	&	0.053	&	6.49	&	0.3	&	0.06	&	0.19	&	0.13	&	1.65	&	20	&	0.0068	&	0.0056	\\
229606609	&	170.33	&	13.71	&	0.989	&	3.14	&	1.17	&	4.88	&	6.43	&	4707	&	-0.06	&	0	&	0.75	&	0.61	&	0.036	&	0.18	&	0.3	&	1.41	&	18	&	0.0025	&	0	\\
229608123	&	360.86	&	22.25	&	1.006	&	3.45	&	1.68	&	3.94	&	1.64	&	5076	&	-0.24	&	0.082	&	31.71	&	1.22	&	0.065	&	0.33	&	0.48	&	1.51	&	25.5	&	0.0062	&	0.11	\\
229739896	&	148.04	&	11.94	&	0.998	&	3.069	&	1.31	&	5.52	&	4.54	&	4753	&	0.02	&	0.035	&	0.99	&	0.45	&	0.071	&	0.13	&	0.53	&	0.39	&	18.5	&	0.0058	&	0.0051	\\
229962295	&	428.46	&	30.61	&	1.014	&	3.88	&	0.92	&	2.6	&	8.9	&	5766	&	0.05	&	-0.019	&	15.55	&	1.01	&	0.007	&	0.34	&	0.45	&	7.08	&	29.5	&	0.0103	&	0.0074	\\
230082687	&	324.51	&	21.44	&	0.996	&	3.43	&	1.39	&	3.82	&	3.91	&	4873	&	0.16	&	0.033	&	16.5	&	0.6	&	0.067	&	0.2	&	0.37	&	0.66	&	24	&	0.0061	&	0.0048	\\
230088835	&	304.02	&	20.21	&	1.002	&	3.42	&	1.47	&	4.03	&	2.83	&	5012	&	-0.07	&	0.039	&	13.53	&	0.82	&	0.065	&	0.11	&	0.08	&	0.18	&	28.5	&	0.0073	&	0.0057	\\
230111524	&	323.54	&	21.89	&	1.001	&	3.42	&	1.24	&	3.62	&	4.71	&	4878	&	-0.19	&	0	&	4	&	0.15	&	0.071	&	0.17	&	0.06	&	0.68	&	29	&	0.013	&	0	\\
232610052	&	222.13	&	17.9	&	0.999	&	3.3	&	0.9	&	3.72	&	14.59	&	4855	&	-0.22	&	0.254	&	15.55	&	1.01	&	0.061	&	0.16	&	0.09	&	4.38	&	18	&	0.0071	&	0.0061	\\
232614734	&	140.06	&	11.1	&	0.993	&	3.1	&	1.58	&	6.19	&	2.45	&	4880	&	0.08	&	0.007	&	10.9	&	0.77	&	0.078	&	0.15	&	0.38	&	0.24	&	14.5	&	0.0063	&	0.005	\\
233047757	&	200.91	&	15.13	&	1.004	&	3.2	&	1.23	&	4.6	&	5.91	&	4736	&	0.11	&	0.084	&	3.03	&	0.49	&	0.061	&	0.15	&	0.33	&	0.78	&	17.5	&	0.0057	&	0.0047	\\
233050540	&	325.08	&	22.15	&	1.008	&	3.45	&	1.26	&	3.58	&	4.02	&	5120	&	-0.33	&	0.084	&	15.55	&	1.01	&	0.055	&	0.16	&	0.04	&	0.55	&	24	&	0.0064	&	0.11	\\
233053435	&	232.36	&	15.84	&	0.998	&	3.24	&	1.68	&	4.97	&	1.91	&	4840	&	0	&	0	&	27.2	&	0.19	&	0.066	&	0.41	&	0.59	&	1.59	&	29.5	&	0.01	&	0	\\
233056745	&	602.38	&	35.68	&	1.005	&	3.69	&	1.14	&	2.53	&	7.03	&	4947	&	-0.02	&	0.043	&	3.05	&	2.03	&	0.045	&	0.1	&	0.1	&	0.52	&	23.5	&	0.0057	&	0.0535	\\
233678984	&	441.12	&	27.83	&	1.016	&	3.57	&	1.25	&	3.06	&	4	&	5214	&	-0.32	&	0.084	&	15.55	&	1.01	&	0.025	&	0.12	&	0.07	&	0.37	&	26.5	&	0.009	&	0.0452	\\
233680706	&	227.93	&	18.73	&	1.007	&	3.31	&	0.8	&	3.46	&	23.05	&	4928	&	-0.1	&	0	&	15.55	&	1.01	&	0.107	&	0.35	&	0.15	&	9.19	&	22.5	&	0.0003	&	0	\\
233683095	&	242.67	&	18.84	&	1.005	&	3.3	&	0.95	&	3.64	&	13.23	&	4896	&	-0.11	&	0.072	&	3.62	&	0.22	&	0.053	&	0.17	&	0.04	&	4.43	&	16.5	&	0.0069	&	0.0058	\\
258389994	&	214.2	&	16.07	&	1	&	3.24	&	1.27	&	4.47	&	4.44	&	4930	&	-0.18	&	0.079	&	1.21	&	0.04	&	0.045	&	0.2	&	0.3	&	1.11	&	20	&	0.0074	&	0.0061	\\
264222470	&	371.77	&	22.42	&	1.002	&	3.46	&	1.77	&	4	&	1.58	&	5001	&	-0.02	&	0.035	&	21.88	&	1.1	&	0.045	&	0.47	&	0.43	&	1.33	&	24.5	&	0.007	&	0.0067	\\
272784360	&	283.42	&	19.9	&	1.01	&	3.35	&	1.24	&	3.83	&	4.16	&	5043	&	-0.36	&	0	&	7.57	&	0.1	&	0.068	&	0.19	&	0.05	&	0.93	&	26.5	&	0.019	&	0	\\
287140025	&	303.35	&	20.8	&	1.005	&	3.4	&	1.29	&	3.78	&	4.3	&	5005	&	-0.11	&	0	&	2.35	&	1.07	&	0.031	&	0.17	&	0.35	&	0.69	&	18.5	&	0.0005	&	0	\\
307985358	&	163.87	&	12.5	&	0.989	&	3.11	&	1.55	&	5.69	&	2.69	&	4782	&	0.1	&	0.026	&	8.51	&	0.44	&	0.075	&	0.27	&	0.66	&	0.67	&	18	&	0.0064	&	0.0055	\\
332503809	&	355.07	&	25.68	&	1.008	&	3.47	&	0.85	&	2.85	&	17.7	&	4899	&	-0.26	&	0.176	&	24.13	&	0.46	&	0.043	&	0.24	&	0.27	&	10.69	&	11	&	0.0074	&	0.0708	\\
356111938	&	415.93	&	26.71	&	1.001	&	3.53	&	1.19	&	3.12	&	6.66	&	4866	&	0.15	&	0.041	&	12.51	&	1.07	&	0.06	&	0.15	&	0.04	&	0.99	&	19.5	&	0.0061	&	0.0048	\\
360019958	&	215.31	&	15.4	&	0.977	&	3.22	&	1.49	&	4.93	&	3.03	&	4563	&	0.04	&	0.021	&	14.46	&	0.14	&	0.007	&	0.32	&	0.34	&	1.13	&	21	&	0.0063	&	0.0051	\\
366490487	&	120.29	&	13.23	&	1.018	&	3.09	&	0.54	&	3.78	&	18.77	&	5499	&	0.18	&	-0.031	&	15.55	&	1.01	&	0.864	&	0.65	&	1.71	&	16.26	&	22.5	&	0.007	&	0.005	\\
441731916	&	339.12	&	25.53	&	1.008	&	3.52	&	0.75	&	2.75	&	14.45	&	4882	&	0.01	&	0.137	&	15.55	&	1.01	&	0.106	&	0.37	&	0.19	&	5.96	&	15.5	&	0.0064	&	0.0142	\\
\enddata

 \vspace{0cm}
 \hspace{1cm}
 \tablecomments{The data table for the entire sample. The listed $\nu_{\text{max}}$, $\Delta\nu$, $log(g)$, $M_{cor}$, and $R_{cor}$ are our calculated seismic values. The $\nu_{\text{max}}$, $\Delta\nu$, and $log(g)$ uncertainties are found from taking the difference between our sample and previous literature and subtracting the previous literature's uncertainties in quadrature. The $M_{cor}$ and $R_{cor}$ uncertainties are found by propagating forward the uncertainties on the $\nu_{\text{max}}$, $\Delta\nu$, $T_{\text{eff}}$, $f_{\nu_{\text{max}}}$, and $f_{\Delta\nu}$. The listed $T_{\text{eff}}$, [M/H], and [$\alpha/M$] values are spectroscopic with $T_{\text{eff}}$ and [M/H] calibrated in \citet{2021ApJ...915...19G} and [$\alpha/M$] directly from APOGEE DR16. For stars with no published APOGEE data, we put a value of 0 for [$\alpha/M$]. The $T_{\text{eff}}$ and [M/H] uncertainties are taken directly from \citet{2021ApJ...915...19G} and the [$\alpha/M$] uncertainties are taken directly from APOGEE DR16. The listed $f_{\Delta\nu}$ values are given by ASFGrid \citep{2016ascl.soft03009S}. The average $f_{\nu_{\text{max}}}$ value is 0.9877 and is found by calibrating our computed luminosity to the {\it Gaia} luminosity calibrated in \citet{2021ApJ...915...19G}. The listed age is the average age from the YREC, MIST, DSEP, and GARSTEC model grids. The age uncertainties are the average observational and theoretical errors from the four stellar model grids added in quadrature. The separate average observational and average theoretical age uncertainties are listed in Table \ref{tab:table2}.}

\end{deluxetable}
\end{longrotatetable}

\onecolumngrid
\newpage
\startlongtable
\begin{deluxetable}{ccccccccccccc}

\tabletypesize{\footnotesize}
\tablewidth{0pt}

 \tablecaption{Age Values and Uncertainties \label{tab:table2}}

 \tablehead{
 \colhead{TIC ID} & \colhead{YREC Ages} & \colhead{MIST Ages} & \colhead{DSEP Ages} & \colhead{GARSTEC Ages} &  \colhead{$\sigma_{\text{Age}_{\text{{the}}}}$} & \colhead{$\sigma_{\text{Age}_{\text{{obs}}}}$}
 }

\startdata
25156036	&	4.43	&	4.35	&	4.35	&	4.35	&	0.04	&	0.26	\\
31506424	&	2.91	&	2.61	&	2.8	&	2.83	&	0.13	&	0.02	\\
38602419	&	10.73	&	9.27	&	10.13	&	10.28	&	0.61	&	4.38	\\
41587424	&	2.39	&	2.16	&	2.39	&	2.36	&	0.11	&	0.18	\\
55270123	&	2.45	&	2.17	&	2.46	&	2.38	&	0.14	&	0.49	\\
141335100	&	9.69	&	9.56	&	8.75	&	9.37	&	0.41	&	1.53	\\
141626634	&	1.85	&	1.7	&	1.89	&	1.74	&	0.09	&	0.13	\\
141757732	&	2.56	&	2.31	&	2.55	&	2.55	&	0.12	&	0.13	\\
142109390	&	2.13	&	1.98	&	2.18	&	2.08	&	0.09	&	3.84	\\
149390710	&	9.6	&	8.32	&	9.21	&	9.31	&	0.55	&	0.25	\\
150030411	&	4.44	&	4.29	&	4.35	&	4.38	&	0.06	&	0.38	\\
150062447	&	22.74	&	19.33	&	21.02	&	21.16	&	1.39	&	15.99	\\
150166759	&	5.03	&	4.78	&	4.93	&	4.97	&	0.11	&	0.77	\\
150393198	&	9.75	&	8.69	&	9.45	&	9.53	&	0.46	&	0.67	\\
150430044	&	1.95	&	1.92	&	1.95	&	1.84	&	0.05	&	1.81	\\
150442152	&	6.04	&	5.62	&	5.77	&	5.94	&	0.19	&	1.57	\\
167342488	&	3.17	&	2.83	&	3.15	&	3.14	&	0.16	&	0.06	\\
167548586	&	5.62	&	5.17	&	5.35	&	5.55	&	0.21	&	1.01	\\
177241155	&	2.69	&	2.4	&	2.69	&	2.67	&	0.14	&	0.04	\\
220457103	&	5.72	&	5.08	&	5.61	&	5.58	&	0.28	&	1.46	\\
257720189	&	7.8	&	7.22	&	7.5	&	7.65	&	0.25	&	1.18	\\
260078030	&	2.86	&	2.49	&	2.89	&	2.73	&	0.18	&	0.65	\\
260655847	&	7.23	&	6.58	&	7.04	&	7.06	&	0.28	&	0.19	\\
271977529	&	4.68	&	4.45	&	4.6	&	4.62	&	0.1	&	0.43	\\
278731442	&	9.14	&	7.97	&	8.84	&	8.93	&	0.51	&	1.53	\\
279055252	&	2.1	&	1.92	&	2.13	&	2.02	&	0.1	&	0.81	\\
279090824	&	2.54	&	2.2	&	2.57	&	2.39	&	0.17	&	0.05	\\
279571746	&	1.95	&	1.79	&	1.98	&	1.89	&	0.08	&	1.77	\\
293271686	&	3.04	&	2.73	&	3.08	&	2.98	&	0.16	&	0.21	\\
300085625	&	8.17	&	7.75	&	7.87	&	7.98	&	0.18	&	1.34	\\
300088321	&	16.98	&	13.84	&	12.93	&	13.1	&	1.89	&	12.02	\\
300161002	&	3.72	&	3.41	&	3.71	&	3.7	&	0.15	&	0.65	\\
300444376	&	1.79	&	1.64	&	1.83	&	1.69	&	0.09	&	1.23	\\
349059821	&	3.42	&	3.03	&	3.34	&	3.39	&	0.18	&	0.06	\\
349374407	&	6.07	&	5.35	&	5.93	&	5.96	&	0.32	&	1.38	\\
349374677	&	4.36	&	3.85	&	4.15	&	4.31	&	0.23	&	0.09	\\
349477633	&	3.63	&	3.24	&	3.57	&	3.6	&	0.18	&	0.29	\\
349973705	&	2.55	&	2.31	&	2.61	&	2.41	&	0.14	&	0.32	\\
350297922	&	7.11	&	6.33	&	6.77	&	6.94	&	0.34	&	0.68	\\
350335258	&	3.74	&	3.36	&	3.63	&	3.7	&	0.17	&	0.85	\\
350343922	&	7.7	&	6.83	&	7.47	&	7.49	&	0.37	&	0.2	\\
350443250	&	2.22	&	1.94	&	2.23	&	2.11	&	0.14	&	0.01	\\
350585437	&	8.34	&	7.64	&	8.16	&	8.14	&	0.3	&	0.99	\\
374858999	&	1.98	&	1.79	&	2	&	1.87	&	0.1	&	0.67	\\
375089600	&	14.35	&	12.48	&	13.44	&	13.69	&	0.78	&	2.98	\\
381976956	&	5.38	&	4.86	&	5.19	&	5.28	&	0.23	&	2.88	\\
382067256	&	1.39	&	1.33	&	1.42	&	1.32	&	0.05	&	0.91	\\
382159212	&	2.5	&	2.26	&	2.54	&	2.36	&	0.13	&	0.72	\\
382302322	&	6	&	5.25	&	5.81	&	5.89	&	0.33	&	1	\\
198383789	&	7.58	&	6.8	&	7.46	&	7.46	&	0.36	&	1.18	\\
219096868	&	2.24	&	2.06	&	2.27	&	2.21	&	0.09	&	0.57	\\
219792330	&	8.19	&	7.2	&	7.97	&	8.02	&	0.44	&	0.67	\\
219801618	&	1.05	&	0.98	&	1.06	&	1.01	&	0.04	&	0.9	\\
219856298	&	5.92	&	5.23	&	5.7	&	5.81	&	0.3	&	0.84	\\
224601836	&	9.62	&	8.57	&	9.22	&	9.32	&	0.44	&	2.66	\\
224605278	&	6.95	&	6.22	&	6.73	&	6.82	&	0.32	&	1.61	\\
229606609	&	6.68	&	6.03	&	6.47	&	6.54	&	0.28	&	1.38	\\
229608123	&	1.71	&	1.51	&	1.72	&	1.63	&	0.09	&	1.51	\\
229739896	&	4.66	&	4.33	&	4.55	&	4.62	&	0.15	&	0.36	\\
229962295	&	11.18	&	8.6	&	7.58	&	8.26	&	1.57	&	6.9	\\
230082687	&	3.97	&	3.78	&	3.97	&	3.92	&	0.09	&	0.65	\\
230088835	&	2.91	&	2.6	&	2.93	&	2.87	&	0.15	&	0.1	\\
230111524	&	4.93	&	4.36	&	4.69	&	4.85	&	0.25	&	0.63	\\
232610052	&	15.41	&	13.37	&	14.83	&	14.76	&	0.87	&	4.3	\\
232614734	&	2.49	&	2.33	&	2.54	&	2.44	&	0.09	&	0.22	\\
233047757	&	6.07	&	5.75	&	5.87	&	5.93	&	0.13	&	0.77	\\
233050540	&	4.23	&	3.71	&	4.01	&	4.14	&	0.23	&	0.5	\\
233053435	&	1.95	&	1.79	&	1.98	&	1.9	&	0.08	&	1.59	\\
233056745	&	7.31	&	6.54	&	7.06	&	7.21	&	0.34	&	0.39	\\
233678984	&	4.22	&	3.7	&	3.98	&	4.09	&	0.22	&	0.3	\\
233680706	&	24.97	&	21.33	&	22.91	&	22.99	&	1.49	&	9.07	\\
233683095	&	14.06	&	12.13	&	13.2	&	13.52	&	0.82	&	4.35	\\
258389994	&	4.63	&	4.11	&	4.42	&	4.57	&	0.23	&	1.09	\\
264222470	&	1.63	&	1.5	&	1.65	&	1.55	&	0.07	&	1.33	\\
272784360	&	4.35	&	3.85	&	4.15	&	4.27	&	0.22	&	0.91	\\
287140025	&	4.49	&	4.01	&	4.28	&	4.43	&	0.21	&	0.65	\\
307985358	&	2.72	&	2.56	&	2.78	&	2.68	&	0.09	&	0.66	\\
332503809	&	19.15	&	15.91	&	17.65	&	18.09	&	1.35	&	10.61	\\
356111938	&	6.86	&	6.54	&	6.55	&	6.68	&	0.15	&	0.97	\\
360019958	&	3.04	&	2.85	&	3.11	&	3.11	&	0.12	&	1.13	\\
366490487	&	23.09	&	17.73	&	16.68	&	17.58	&	2.92	&	16	\\
441731916	&	14.27	&	11.19	&	16.32	&	16.02	&	2.35	&	5.47	\\
\enddata

 \vspace{0cm}
 \tablecomments{The ages for the YREC, MIST, DSEP, and GARSTEC stellar model grids for the entire sample. The $Age_{the}$ error is the average theoretical error given by \texttt{Kiauhoku} between the four model grids. The $Age_{obs}$ error is the average observational error calculated by propagating forward the uncertainty on the mass and metallicity. The units for all ages and the error are in Gyr.}

\end{deluxetable}

\end{document}